\documentclass[11pt]{article}
\usepackage[dvips]{graphicx}
\usepackage{amssymb,amsmath,color}
\usepackage{url}

\usepackage[mathcal]{eucal}

\usepackage{changes}

\DeclareMathAlphabet\mathbfcal{OMS}{cmsy}{b}{n}

\usepackage[colorlinks=true, allcolors=blue, pagebackref]{hyperref}       
\renewcommand*{\backrefalt}[4]{%
    \ifcase #1 \footnotesize{(not cited)}%
    \or        \footnotesize{(cited on page~#2)}%
    \else      \footnotesize{(cited on pages~#2)}%
    \fi}

\newcommand{\BEAS}{\begin{eqnarray*}}
\newcommand{\EEAS}{\end{eqnarray*}}
\newcommand{\BEA}{\begin{eqnarray}}
\newcommand{\EEA}{\end{eqnarray}}
\newcommand{\BEQ}{\begin{equation}}
\newcommand{\EEQ}{\end{equation}}
\newcommand{\BIT}{\begin{itemize}}
\newcommand{\EIT}{\end{itemize}}
\newcommand{\BNUM}{\begin{enumerate}}
\newcommand{\ENUM}{\end{enumerate}}
\newcommand{\BA}{\begin{array}}
\newcommand{\EA}{\end{array}}

\newcommand{\tr}{\mathop{ \rm tr}}

\newcommand{\idm}{I}
\newcommand{\rb}{\mathbb{{R}}}
\newcommand{\cb}{\mathbb{C}}
\newcommand{\zb}{\mathbb{Z}}

\newcommand{\ds}{\displaystyle }
\newcommand{\BlackBox}{\rule{1.5ex}{1.5ex}}  

\newtheorem{lemma}{Lemma}

\newcommand{\mysec}[1]{Section~\ref{sec:#1}}
\newcommand{\eq}[1]{Eq.~(\ref{eq:#1})}
\newcommand{\myfig}[1]{Figure~\ref{fig:#1}}

 \def \ds { \displaystyle}

\def \E{{\mathbb E}}

\def \H{{\mathbb{H}}}

\def \X{{\mathcal X}}

\title{Sum-of-Squares Relaxations
 \\ for Information Theory and Variational Inference}

\author{Francis Bach}

\date{\today}

\begin{document}
\maketitle

\begin{abstract}
We consider extensions of the  Shannon relative entropy, referred to as $f$-divergences.
Three classical related computational problems are typically associated with these divergences: (a) estimation  from moments, (b) computing normalizing integrals, and (c) variational inference in probabilistic models. These problems are related to one another through convex duality, and 
  for all them, there are many applications throughout data science, and we aim for computationally tractable  approximation algorithms that preserve properties of the original problem such as potential convexity or monotonicity. In order to achieve this, we derive a sequence of convex relaxations for computing these divergences from non-centered covariance matrices associated with a given feature vector: starting from the typically non-tractable optimal lower-bound, we consider an additional relaxation based on ``sums-of-squares'', which is is now computable in polynomial time as a semidefinite program. We also provide computationally more efficient relaxations based on spectral information divergences from quantum information theory. For all of the tasks above, beyond proposing new relaxations, we derive tractable convex optimization algorithms, and we present illustrations on multivariate trigonometric polynomials and functions on the Boolean hypercube.
\end{abstract}

 \section{Introduction}
 
Tools from information theory are ubiquitous in data science. Starting with the notion of Shannon entropy, other notions have emerged, in particular $f$-divergences~\cite{csiszar1967information,ali1966general}, which are defined as 
\BEQ
\label{eq:Dpq}
D(p\|q) = \int_\X f\Big( \frac{dp}{dq}(x) \Big) dq(x),
\EEQ
 where $p$ and $q$ are two finite positive measures on an arbitrary  set $\X$, $\frac{dp}{dq}$ is the density of $p$ with respect to~$q$, and $f: \rb_+^\ast \to \rb$ is a convex function.\footnote{{Note that our notation $D(p\|q)$ ignores the dependence in $f$.}} A classical example is $f(t) = t \log t - t + 1$, where $D(p\|q)$ is the usual Kullback-Leibler divergence, associated with Shannon information theory~\cite{cover1999elements}, which we will use as a running example.
 
 These divergences have been used in many areas in machine learning, signal processing or statistics, such as within message passing and variational inference \cite{minka2005divergence}, {concentration inequalities~\cite{boucheron2013concentration}}, PAC-Bayes analysis~\cite{picard2022change}, independent component analysis~\cite{cardoso2003dependence}, information theory~\cite{polyanskiy2022information}, differential privacy~\cite{mironov2017renyi}, design of surrogate losses for classification~\cite{nguyen2009surrogate}, and optimization~\cite{ben1987penalty}. 
 We review $f$-divergences and their basic properties in  \mysec{f-review}, see~\cite{liese2008f,liese2006divergences,sason2018f} for a more complete treatment.

Two classical related computational problems are typically associated with $f$-divergences, which have to be estimated  or optimized in some way, a task that can become difficult in multivariate settings. For all them, there are many applications throughout data science, and we aim for computationally tractable algorithms that preserve properties of the original problem (such as potential convexity or monotonicity). 

 \BIT

\item[(1)] \textbf{Estimation of  divergences from moments:} Given some function $T$ from $\X$ to some vector space, the goal is to estimate $D(p\|q)$ {defined in \eq{Dpq}} only from the knowledge
of the integrals $  \int_\X T(x) dp(x)$ and  $  \int_\X T(x) dq(x)$. Our aim in this paper is to estimate $D(p\|q)$ from below, and to obtain the largest possible lower bound. We focus on particular functions $T$ of the form $T(x) = \varphi(x) \varphi(x)^\ast$, where $\varphi: \X \to \cb^d$ is some complex-valued feature map, and where $M^\ast$ denotes the conjugate transpose of the matrix $M$. Thus, in our particular situation, $T$ takes values in the set $\mathbb{H}_d^+$ of positive semi-definite Hermitian matrices of size $d \times d$. {This choice of the feature map $T$ as a rank-one Hermitian matrix is not a limitation in many instances, such as with polynomials (as monomials can be arranged in Hankel matrices) and is key to our methodological developments.}

For this particular form of moments as non-centered covariance matrices, we first provide in \mysec{exact} a characterization of the tightest such lower bound. This formulation involves the maximization over~$\X$ of {quadratic forms in $\varphi$, that is,} functions of the form $x \mapsto \varphi(x)^* M \varphi(x)$, where $M \in \mathbb{H}_d$ {(the set of Hermitian matrices of size $d \times d$)}.

Our first contribution is to replace the exact maximization of such quadratic forms of $\varphi(x)$ by ``sum-of-squares'' relaxations, that is,  relaxations based on semi-definite programming and the representation of non-negative functions as positive-semidefinite quadratic forms in  $\varphi(x)$~\cite{lasserre2010moments,parrilo2003semidefinite} (see review in \mysec{sos}). This relaxation is developed in \mysec{sos-relax} and allows to bring to bear the well-developed area of sum-of-squares optimization with its computational tools and extensive analyses. We also provide in \mysec{qt-relax} a further relaxation which is based on information divergences from quantum information theory (which are reviewed in \mysec{qit}). 

Note that a related interesting task is to estimate estimation divergences directly from samples~\cite{nguyen2010estimating,rubenstein2019practical}. We could use our algorithms with increasingly large feature vectors and use empirical estimates, but a detailed analysis is left for future research.

\item[(2)] \textbf{Variational inference in probabilistic models:}
One classical inference task in probabilistic modeling 
{(see~\cite{wainwright2008graphical,murphy2012machine} and references therein)}
is to compute moments of some distributions from which we know the density. In our  context of $f$-divergences, we consider a density (with respect to {some positive measure} $q$) proportional to $(f^\ast)'( h(x) - \rho )$, {where $f^\ast$ is the Fenchel conjugate of $f$}, $h: \X \to \rb$ is an arbitrary function, and $\rho \in \rb$ is a normalizing constant making sure that we obtain a probability distribution. As shown in \mysec{fpart}, this density happens to be exactly the maximizer in  
$$ c_q(h) = \sup_{p\  {\rm probability \ measure \ on}\  \X}\  \int_\X h(x) dp(x) - D(p\|q).$$ The optimal quantity  $c_q(h)$ is referred to as the $f$-partition-function, and for $f(t) = t \log t - t + 1$, we recover the usual log-partition function, and densities proportional to $e^{h(x)}$.

When we restrict $h$ to be a quadratic form  in $\varphi(x)$, that is of the form $\varphi(x)^\ast H \varphi(x)$ for some $H \in \mathbb{H}_d$, then, (a)  {we can replace $D(p\|q)$ by the lower-bound we just defined above, and obtain a computable \emph{upper}-bound of $c_q(h)$}, and (b)
the gradient with respect to $H$ of the $f$-partition function ends up being exactly the moment of $T(x) = \varphi(x) \varphi(x)^\ast \in \H_d$ for the desired distribution. {This relaxation is presented in  \mysec{partition}, and can be extended to the task of computing integrals of the form $ \int_{\X} f^\ast ( h(x)) dq(x)$ (see Appendix~\ref{app:computingintegrals}).}

{For well-chosen feature vectors, e.g., polynomials on $\{-1,1\}^d$, log-densities that can be expressed as quadratic forms cover a wide set of Markov random field models in statistical modeling~\cite{lauritzen1996graphical} and image processing~\cite{blake2011markov}.  For these models, exact moment estimation and log-partition estimations are key computational tasks that are intractable, even in moderate dimensions, hence the need for approximations. The formulations presented in this paper follow a line of work based on tractable convex relaxations, typically based on linear programming, with few examples using the more powerful semi-definite programming framework that we further develop (see \cite{wainwright2008graphical} for a thorough introduction).}

\EIT

\paragraph{Contributions.}
In this paper, we first derive a sequence of three convex formulations of $f$-divergences based on covariance matrices. Starting from the typically non-tractable optimal lower-bound, we consider an additional relaxation based on ``sums-of-squares'', which is now computable in polynomial time as a semidefinite program, as well as further computationally more efficient relaxations based on spectral information divergences from quantum information theory. For all of the tasks above, beyond proposing new relaxations, we derive tractable algorithms based on convex optimization, and we present illustrations on multivariate trigonometric polynomials and functions on the Boolean hypercube. We then extend these bounds by duality to lower bounds on partition functions.

{Since these contributions involve three traditionally separate domains, we start by a review of those, that is, $f$-divergences and associated variational formulations in \mysec{f-review} (where we propose a new one more adapted to our purposes), quantum information divergences in \mysec{qit}, and finally sum-of-squares relaxations in \mysec{sos}.}

 \section{Review of $f$-divergences}
\label{sec:f-review}

We consider $f$-divergences, with $f: \rb_+^\ast \to \rb_+$ is a convex function, {where $\rb_+^\ast$ denotes the set of strictly positive real numbers}. We assume that $f$ is strictly convex and differentiable, so that the Fenchel conjugate~$f^\ast$ is differentiable and non-decreasing, with $(f^\ast)'(u) \geqslant 0$ for $u$ in the domain of $f^\ast$. Moreover, we assume that $f(1)=0$, and thus $1$ is the minimizer of $f$, leading to $f'(1)=0$ and $(f^\ast)'(0) = 1$. Moreover, we then have $f^\ast(0) = 0$. Our running example is $f(t) = t \log t - t + 1$ with $f^\ast(u) = e^u -1$ (see more examples below).

On the set $\X$ (which we only assume to be equipped with a topology, and compact),  we consider several sets of finite Borel measures: $\mathcal{M}_+(\X)$ the set of finite \emph{positive} measures on~$\X$, 
$\mathcal{M}(\X)$ the set of finite \emph{signed} measures on~$\X$, and $\mathcal{P}(\X)$ the set of  \emph{probability} measures on~$\X$ (that is, {finite} positive measures in $\mathcal{M}_+(\X)$ that integrate to one).

For two finite positive measures $p,q$ in  $\mathcal{M}_+(\X)$, we can  define
$$
D(p\|q) = \int_\X f\Big( \frac{dp}{dq}(x) \Big) dq(x),
$$
for all non-negative measures (possibly non normalized), assuming that the density $\frac{dp}{dq}(x) $ exists for all $x \in \X$ and that the integral is finite.   We now review several properties and examples, see~\cite{csiszar1967information,sason2018f,picard2022change} for more results.

\paragraph{Classical properties.}
Given our assumption that $1$ is a global minimizer of $f$, $f(1) = 0$, and $f$ is strictly convex, we have $D(p\|q) \geqslant 0$ with equality if and only if $p=q$. Moreover, $D(p\|q)$ is jointly convex in $p$ and $q$.

\paragraph{Examples.}
We have the following classical examples, with the usual ``reversion'' of $f$-divergences: if we define $g(t) = t f(1/t)$, swapping $p$ and $q$ in $D(p\|q)$ is equivalent to replacing $f$ by $g$ (for $\alpha$-divergences below, this corresponds to replacing $\alpha$ by $ 1- \alpha$). All of the approximations that we consider in this paper will satisfy this reversibility: swapping $p$ and $q$ (and later moment matrices $A$ and $B$) is equivalent to replacing $f$ by~$g$.

Note that the total variation case, where $f(t) = |t-1|$  is excluded from most developments because it is neither differentiable nor strictly convex (nor operator convex, {as defined in \mysec{opcvx}}), but many results (except the quantum ones) would apply as well. We normalize all functions $f$ so that $f''(1)=1$. See table and plots below.

\begin{center}
\begin{tabular}{|l|l|l|l|l|}
\hline
Divergence & $f(t)$ & $f^\ast(u)$  & $(f^\ast)'(u)$  \\
\hline
$\alpha$-R\'enyi & $\textcolor{white}{\Big|}  \!\!\!\!\frac{1}{\alpha(\alpha-1)} \big[   t^\alpha - \alpha t + (\alpha\!-\!1) \big]\!\!\!$
& $ \!\!\frac{1}{\alpha} \big[ \! -\! 1 \!+\! ( 1\! +\! (\alpha\!-\!1) u )^{\alpha/(\alpha-1)} \big]\!\!$  & $\!(1\! +\! (\alpha\!-\!1) u)^{1/(\alpha-1)}\!\!$ \\
Kullback-Leibler, $\alpha\!=\! 1$ & $ t \log t -t + 1 $  & $ e^{u}-1$   & $e^u$\\
Reverse KL, $\alpha\!=\! 0$ & $- \log t + t - 1$ & $  - \log(1-u)$  & $\frac{1}{1-u} $ \\
squared Hellinger, $\alpha\!=\! \frac{1}{2}\!\!$ &  $2 (\sqrt{t}-1)^2$  & $\frac{u}{1-u/2}$ & $\frac{1}{( 1-u/2)^2} $\\
Pearson $\chi^2$, $\alpha\!=\!2$ & $\textcolor{white}{\Big|} \frac{1}{2} (t-1)^2$ & $ \frac{1}{2} (u+1)_+^2 - \frac{1}{2}$ & $(u+1)_+$\\
Reverse Pearson, $\alpha\!=\!-\!\,1\!\!$ & $\frac{1}{2} \big( \frac{1}{t} + t \big) - 1$ & $1- \sqrt{1-2u}$ & $\frac{1}{\sqrt{1-2u}}$\\
Le Cam & $ \frac{(t-1)^2}{t+1} $ & $2-u - 2\sqrt{1-2u}$  &  $ \frac{2}{\sqrt{1-2u}}-1$ \\
 Jensen-Shannon & $  2 t \log \frac{2t}{t+1} +  2\log \frac{2}{t+1}$ & $-2 \log(2-e^{u/2})$  &   $ \frac{1}{2\exp(-u/2)-1}$ \\
\hline
\end{tabular}
\end{center}

\begin{center}
\hspace*{-2cm} \includegraphics[width=21cm]{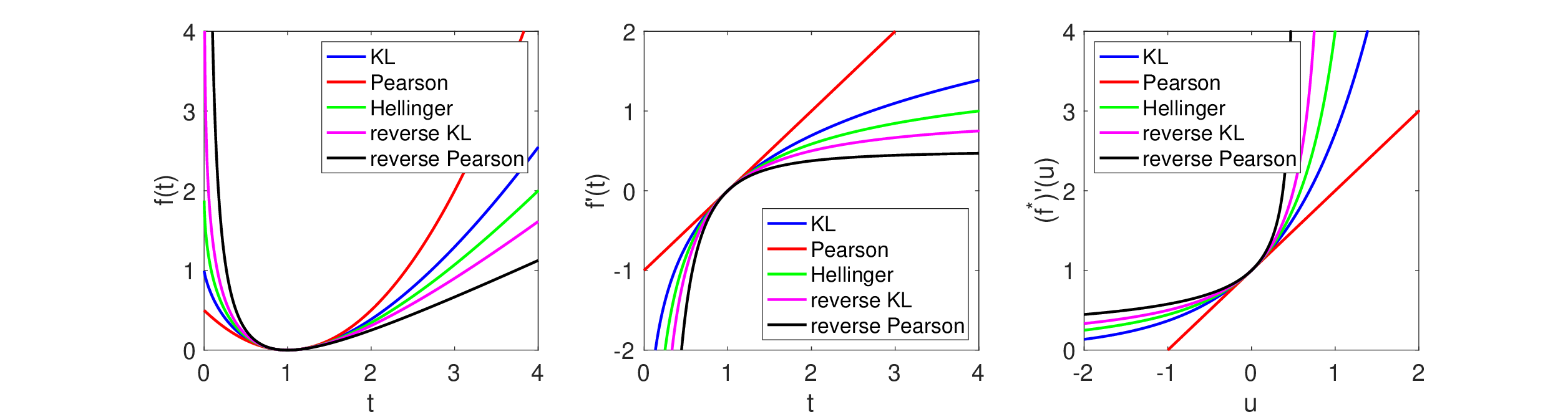}
\end{center}

\subsection{Variational representations} 
\label{sec:varf}

The $f$-divergence  has a variational representation obtained from  the Fenchel conjugate of perspective functions~\cite{rockafellar2015convex}.
Indeed, the function  $  (p,q) \mapsto q f\big( \frac{p}{q} \big)$ defined on $\rb_+ \times \rb_+^\ast$ is referred to as the \emph{perspective function} of $f$, {is convex,} and has the variational representation for $p \in \rb_+$ and $q \in \rb_+^\ast$:
\BEQ
\label{eq:persp}
q f\Big( \frac{p}{q} \Big) = \sup_{ v,w \in \rb} \ vp + w q \ \mbox{ such that }  \ \forall r \geqslant 0, \  r  v + w \leqslant f(r),
\EEQ
where for $p$ and $q \in \rb_+^\ast$, {the maximizer $w$ is $\inf_{r \geqslant 0} f(r) - rv = -f^\ast(v)$, and thus the optimal value of $v$ is the supremum of $vp - qf^\ast(v)$, leading to optimal values} $v^\ast = f' \big( \frac{p}{q} \big)$, and $w^\ast = - f^\ast(v^\ast) = f \big( \frac{p}{q} \big) -    \frac{p}{q} f' \big( \frac{p}{q} \big)$. {Depending on the behavior of $f$ at $0$ and $+\infty$, we can extend the perspective function to $\rb_+\times \rb_+$.}

Following~\cite{matsumoto2015new}, for $p,q \in  \mathcal{M}_+(\X)$, {applying \eq{persp} to densities}, this leads to a variational representation of $D(p\|q)$ as the supremum of linear functions of the measures $p$ and $q$ {(with functions $v$ and $w$ that are measurable and bounded)}:
\BEQ
\label{eq:f-var}
 D(p\|q) =   \sup_{ v,w : \X \to \rb } \ 
  \int_\X v(x) dp(x)  + \int_\X w(x) dq(x)  \  \mbox{ such that }  \ \forall x \in \X, \ \forall r \geqslant 0, \  r v(x)  + w(x) \leqslant f(r).
\EEQ
The optimal functions $w$ and $v$ are such that  
$v(x) =  f'\big( \frac{dp}{dq}(x) \big)$, and $w(x) =  f\big( \frac{dp}{dq}(x) \big)  - \frac{dp}{dq}(x)    f'\big( \frac{dp}{dq}(x) \big) = -f^\ast(v(x))$. {We can also use the function $g$ defined above through $g(t) = t f(1/t)$, to get $w(t) =  g'\big( \frac{dq}{dp}(x) \big)$.}

Note that in this representation, the non-negativity of the measures $p$ and $q$ is automatically satisfied (the value of the optimization problem in \eq{f-var} is infinite otherwise). Optimizing with respect to $w(x)$ in closed form as above then leads to the representation {of $D(p\|q)$} from~\cite{broniatowski2006minimization,nguyen2010estimating} as the supremum with respect to $v: \X \to \rb$ of  $  \int_\X v(x) dp(x)  - \int_\X f^\ast(v(x)) dq(x)$. 

\paragraph{New variational unconstrained formulation.} {In this paper, we will need a novel \emph{unconstrained} variational formulation similar to \eq{f-var}. To this effect, we introduce the function $F: \rb^2 \to \rb$ defined as
\BEQ
\label{eq:defF}
F(v,w) = \sup_{r \geqslant 0} \frac{ r  v + w - f(r)}{r+1}.
\EEQ
The function $F$ is convex as a supremum of affine functions and given our assumption on $f$ that $f(t)\geqslant f(1) = 0$ for all $t >0$, we have $\frac{v+w}{2} \leqslant F(v,w) \leqslant \max\{v,w\}$ (which also shows it has full domain). In addition all subgradients of $F$ are in the simplex in $\rb^2$. Moreover, 
for any constant $u \in \rb$, $F(v-u,w-u) = F(v,w)- u$, and $F(w,v) \leqslant 0$ if and only if for all $r\geqslant 0$, $ r  v + w - f(r) \leqslant 0$.}

{Thus, starting from \eq{persp}, we have, for $p,q \in \rb_+^\ast$:
\BEA
 \notag  q f\Big( \frac{p}{q} \Big) &=& \sup_{ v,w \in \rb} \ vp + w q \ \mbox{ such that } F(v,w) \leqslant 0 \\
 \notag &=& \sup_{ u,v,w \in \rb} \ (v-u)p + (w-u) q \ \mbox{ such that } F(v-u,w-u) \leqslant 0, \mbox{ by adding an extra variable } u,\\
 \notag &=& \sup_{ u,v,w \in \rb} \ vp+wq - (p+q) u \ \mbox{ such that } F(v,w) \leqslant u, \mbox{ using properties of } F,  \\
 \label{eq:AAA} &=&  \sup_{ v,w \in \rb} \ vp + w q - (p+q) F(v,w) \mbox{ since the optimal } u \mbox{ is } F(v,w).
\EEA
Using the same technique as above and replacing $w(x)$ and $v(x)$ by $w(x)-u$ and $v(x)-u)$ in \eq{f-var} , we can optimize with respect to $u$ and thus obtain:
as
\BEQ
\label{eq:AD} D(p\|q) =   \sup_{ v,w : \X \to \rb } \ 
  \int_\X v(x) dp(x)  + \int_\X w(x) dq(x) 
  - \Big(
   \int_\X  dp(x)  + \int_\X  dq(x) 
  \Big)  \sup_{x \in \X} F\big(v(x),w(x)\big).
\EEQ
\eq{AD} above will be crucial when estimating $f$-divergences of partition functions because it leads to \emph{unconstrained} optimization problems, while \eq{f-var} was constrained.}

\paragraph{Computing $F$.} {In our algorithms in later sections, we will need to compute $F$ for any $(v,w)\in \rb^2$, that is, solve for $r \geqslant 0$ in \eq{defF}. This is a one-dimensional root-finding problem for which Newton method can be used with quadratic convergence, and thus with few iterations. See Appendix~\ref{app:newton} for details for the function $f(t) = t \log t - t + 1$.}

\paragraph{Variational formulations as infimum.} {Through convex duality, we can derive variational formulations of $D(p\| q)$ as minimization problems rather than maximization problems like in \eq{f-var} and \eq{AD}. Since this is not crucial to our developments, this is presented in Appendix~\ref{app:varmin} for all our formulations.
}

\section{Review of quantum information theory}
\label{sec:qit}

{In order to define quantum information divergences, we first need to introduce operator convexity.}

\subsection{Operator convexity}
\label{sec:opcvx}
{All the examples} of convex functions proposed in \mysec{f-review} also happen to be ``operator convex'', meaning that for two positive semi-definite Hermitian matrices $A$, $B$, and any $\lambda \in [0,1]$,
$$
f( \lambda A + (1-\lambda) B) \preccurlyeq \lambda f(A) + (1-\lambda) f(B),
$$
where $\preccurlyeq$ defines the L\"owner order between Hermitian matrices ($A \preccurlyeq B$ if and only if $B-A$ is positive semi-definite), and $f(A)$ is the spectral function defined as $f(A) = \sum_{i=1}^d f(\lambda_i) u_i u_i^\ast$ when $A = \sum_{i=1}^d \lambda_i u_i u_i^\ast$ is an eigenvalue decomposition of $A$.

A classical necessary and sufficient condition for $f$ being operator-convex is the existence of  a representation of $f$ as, {for $f$ additionally satisfying $f(1)=f'(1)=0$}:
\BEQ
\label{eq:opcvx-rep}
 f(t) =  \beta ( t-1)^2 + (t-1)^2\int_0^{+\infty} \frac{1}{\lambda + t} d\nu(\lambda),
\EEQ
 for some   $\beta \in \rb_+$ and  a  positive measure $\nu$ on $\rb_+$~\cite{bhatia2013matrix}.  When the function $f$ is extendable to an analytic function on $\cb$, then the measure $ \nu$   can be obtained from the Stieltjes inversion formula~\cite{widder1938stieltjes}, as the limit of the measure with density $ \frac{1}{\pi} {\rm Im} \Big( \frac{f(-\lambda - it)}{(\lambda + it + 1)^2} \Big) - \beta$ when $t \to 0^+$. In Appendix~\ref{app:op-cvx}, we provide this decomposition for the examples from the beginning of \mysec{f-review}.  
 
 Operator convexity is crucial for the quantum information divergences that we now consider.
 
\subsection{Quantum information divergences}
\label{sec:qt-review}
We consider two Hermitian positive semi-definite matrices $A$ and $B$ in $\mathbb{H}_d^+$. If $A$ and $B$ commute, then they are jointly diagonalizable, and we can naturally define a divergence as
$$
 \sum_{i=1}^d \lambda_i(B) f \Big( \frac{ \lambda_i(A)}{\lambda_i(B) } \Big),
$$
where $\lambda_i(A)$ and $\lambda_i(B)$ are the corresponding {non-negative} eigenvalues of $A$ and $B$ (with the same eigenvectors). When $A$ and $B$ do not commute, there are several notions of $f$-information divergences that reduce to the formula above when matrices commute~\cite{tomamichel2015quantum}. Among the several candidates from quantum information theory~\cite{matsumoto2015new,fawzi2021defining,hiai2017different}, two are particularly interesting in our context.

The so-called \emph{maximal divergence}  is equal to
$$
\tilde{D}^{\rm QT}_{\max}(A\|B) = \tr \big[ B^{1/2} f( B^{-1/2} A B^{-1/2}) B^{1/2} \big] =  \tr \big[ B  f( B^{-1/2} A B^{-1/2})  \big] ,
$$
while the \emph{standard divergence} is equal to:
\BEQ
\label{eq:sta}
\tilde{D}^{\rm  QT}_{\rm standard}(A\|B)  =   {\rm vec}(B^{1/2})^\ast f(A \otimes B^{-1})   {\rm vec}(B^{1/2}),
\EEQ
 
with the usual Kronecker product notation between matrices and ${\rm vec}(M)$ the column vector obtained by stacking the columns of $M$~\cite{golub83matrix}. It is equal to
$$
\sum_{i,j=1}^d \lambda_i f \Big( \frac{\mu_j}{\lambda_i} \Big) | u_i^\ast v_j | ^2,
$$
where $A = \sum_{j=1}^d \mu_j v_j v_j^\ast$ and $B = \sum_{i=1}^d \lambda_i u_i u_i^\ast$ are  eigenvalue decompositions of $A$ and $B$. Both are jointly convex and equal to zero if and only if $A=B$.

An important feature of these divergences is that they can both be computed in closed form from spectral decompositions. This will give a strong computational advantage for the relaxations that are based on these.

\paragraph{Examples of the standard divergence.} We have the following classical examples below, with simpler formulas than \eq{sta}, {where we recover the von Neumann relative entropy and classical matrix formulations of the R\'enyi entropies.}

\begin{center}
\begin{tabular}{|l|l|l|}
\hline
Divergence & $f(t)$ &  $\textcolor{white}{\Big|} \tilde{D}^{\rm QT}_{\rm standard}(A\|B)$   \\
\hline
$\alpha$-R\'enyi & $\textcolor{white}{\Big|}  \!\!\!\frac{1}{\alpha(\alpha-1)} \big[   t^\alpha - \alpha t + (\alpha\!-\!1) \big]\!\!$
& $\textcolor{white}{\Big|}  \!\!\!\frac{1}{\alpha(\alpha-1)} \big[  \tr [ B^{1-\alpha} A^\alpha ]  - \alpha \tr[A]+ (\alpha\!-\!1) \tr[B]\big]\!\!$ \\
Kullback-Leibler, $\alpha=1$ & $ t \log t -t + 1 $  & $\tr\big[ A \log A - A \log B \big]$ \\
squared Hellinger, $\alpha=\frac{1}{2}\!\!$ &    $2 (\sqrt{t}-1)^2$ & $2 \tr A + 2 \tr B - 4 \tr \big[ A^{1/2} B^{1/2} \big] $ \\
Pearson $\chi^2$, $\alpha=2$ & $\textcolor{white}{\Big|} \frac{1}{2} (t-1)^2$ & $\frac{1}{2} \tr \big[ B^{-1} ( B-A)^2 \big]$\\
\hline
\end{tabular}
\end{center}

From the representation of operator convex functions in \eq{opcvx-rep}, we can infer properties of these divergences from the particular example $f(t) = \frac{(t-1)^2}{\lambda + t}$ for $\lambda >0$, for which we have
$$\tilde{D}^{\rm QT}_{\rm standard}(A\|B)  =   {\rm vec}(A-B)^\ast ( A \otimes \idm + \lambda \cdot B \otimes I )^{-1} {\rm vec} ( A-B),$$
and
$$
\tilde{D}^{\rm QT}_{\max}(A\|B)  =  \tr \big[ (A-B) (A+\lambda B)^{-1} (A-B) \big].
$$
This shows immediately that the two quantum divergences are jointly convex in $A$ and $B$. A less direct property is that for all $A$ and $B$ (see proof in~\cite[Prop.~4.1]{hiai2017different}):
$$ \tilde{D}^{\rm QT}_{\rm standard}(A\|B) \leqslant\tilde{D}^{\rm QT}_{\max}(A\|B). $$
Thus, in our context of lower bounds on the regular $f$-divergence $D(p\|q)$, we get a tighter result with~$ \tilde{D}^{\rm QT}_{\max}$, and a strict improvement over~\cite{bach2022information} which uses $ \tilde{D}^{\rm QT}_{\rm standard}$ in the same context (but only for the Kullback-Leibler divergence). Moreover, the key property outlined by \cite{bach2022information} that 
$\tilde{D}^{\rm QT}_{\rm standard}(A\|B) \leqslant D(p\|q)$ for $A = \int_\X \varphi(x)\varphi(x)^\ast dp(x)$ and $B = \int_\X \varphi(x)\varphi(x)^\ast dq(x)$ as soon as for all $x \in \X$, $\| \varphi(x)\|^2 \leqslant 1$, is preserved for $\tilde{D}^{\rm QT}_{\max}$.
{In fact, in this paper, we derive a sequence of lower bounds that are all improvements on~\cite{bach2022information}. See Table~\ref{tab:summary} in \mysec{conclusion} for a summary.}

While the Fenchel conjugate of $\tilde{D}^{\rm QT}_{\rm standard}(A\|B) $ with respect to $A$ can be computed in closed form in most cases, this is not the case for $\tilde{D}^{\rm QT}_{\rm max}(A\|B) $. Thus, some of the algorithms from~\cite{bach2022information} cannot be extended and we need to derive new ones based on the unconstrained variational formulation presented in \mysec{varf}.

\paragraph{Special case of von Neumann relative entropy.} When $f(t) = t \log t - t + 1$, for the standard divergence $\tilde{D}^{\rm QT}_{\rm standard}(A\|B) $, we get $\tr\big[ A \log A - A \log B \big]$, which is the Bregman divergence associated with the von Neumann entropy $A \mapsto \tr [ A \log A]$. Note that this is different from seeing that $A$ and $B$ are covariance matrices, and considering the Kullback-Leibler between zero-mean Gaussian distributions  with these covariance matrices (which would lead to $\frac{1}{2}\tr [ A B^{-1} ] - \frac{1}{2} \log \det [ A B^{-1}] - \frac{d}{2}$). For an approach linking semi-definite programming and Gaussian entropies, see~\cite{jordan2003semidefinite}.

\section{Review of sum-of-squares relaxations}
\label{sec:sos}
In this section, we assume that $\varphi$ is {continuous on $\X$, and thus bounded since $\X$ is assumed compact}. {To make the results simpler, we assume that features are normalized to unit norm, that is, $\forall x \in \X$, $\| \varphi(x)\|=1$, for the standard Hermitian norm.} We consider the task of computing
\BEQ
\label{eq:gamma}
\Gamma(M) = \max_{x \in \X} \ \varphi(x)^\ast M \varphi(x),
\EEQ
for some matrix $M \in \mathbb{H}_d$. Since $\varphi$ is bounded, $\Gamma$ is a positively homogeneous everywhere finite convex function on $\mathbb{H}_d$. We now introduce necessary tools and notations for presenting sum-of-squares (SOS) relaxations~\cite{lasserre2010moments,parrilo2003semidefinite}.

 Let  $\mathcal{K}$ be the closure of the convex hull of all $\varphi(x)\varphi(x)^\ast$, $x \in \X$,  $\mathcal{C}$ the closure of its conic hull, and $\mathcal{V}$ its linear span. By construction, we have $\mathcal{K} \subset \mathcal{C} \subset \mathcal{V}$, and, since we have assumed that for all $x \in \X$, $\| \varphi(x)\|$=1, we have:
 $$
\Sigma \in \mathcal{K} \ \ \Leftrightarrow \ \ \Sigma \in \mathcal{C} \mbox{ and } \tr [ \Sigma  ] = 1.
$$
{We make the extra assumption that $\mathcal{V}$ contains a positive definite matrix (this will most often be the identity matrix in examples in \mysec{examples}).}

By definition of $\Gamma$ in \eq{gamma}, and {since maximizing linear functions of $\varphi(x) \varphi(x)^\ast$ with respect to $x \in \X$ leads to the same value as maximizing over its convex hull}, we get:
\BEAS
 \Gamma(M) = \max_{x \in \X} \ \  \varphi(x)^\ast M \varphi(x)  = \max_{ \Sigma \in \mathcal{K}} \ \ \tr [ \Sigma M],
 \EEAS
that is, the function $\Gamma$ is the support function of {the convex set} $\mathcal{K}$. Moreover, using our notations for finite measures from \mysec{f-review}, we have $\mathcal{V} =  \big\{ \int_\X \varphi(x) \varphi(x)^\ast dp(x), \ p \in \mathcal{M}(\X)\big\}$, $\mathcal{C} =  \big\{ \int_\X \varphi(x) \varphi(x)^\ast dp(x), \ p \in \mathcal{M}_+(\X)\big\}$ and $\mathcal{K} =   \big\{ \int_\X \varphi(x) \varphi(x)^\ast dp(x), \ p \in \mathcal{P}(\X)\big\}$.

\subsection{Outer approximations of convex hulls}
In order to obtain an \emph{upper-bound} on $\Gamma(M)$ defined in \eq{gamma}, we will look for \emph{outer} approximations of the set $\mathcal{K}$. By construction, the convex hull is included in the affine hull, that is,  if $\Sigma \in \mathcal{K}$, then $\tr [ \Sigma   ] = 1$ and $\Sigma \in \mathcal{V}$. The extra condition we will use in this paper follows~\cite{lasserre2010moments,parrilo2003semidefinite}, and is simply that $\Sigma$ is positive semi-definite, which is a direct consequence of $\varphi(x) \varphi(x)^\ast \in \H_d^+$ for all $x \in \X$.

  We thus consider outer approximations of $\mathcal{K} = \mathcal{C} \cap \{ \Sigma,\  \tr [ \Sigma   ]  = 1\}$, through the outer approximation of $\mathcal{C}$ as 
$\widehat{\mathcal{C}} = \mathcal{V} \cap \mathbb{H}_d^+$, which corresponds to 
$\widehat{\mathcal{K}} =  \mathcal{V}  \cap \mathbb{H}_d^+ \cap \{ \Sigma, \ \tr [  \Sigma] = 1 \}  $, with $\mathbb{H}_d^+$ the set of PSD Hermitian matrices. 
This leads to our approximation of $\Gamma(M)$ as:
\BEA
\label{eq:gammaaa}\widehat{\Gamma}(M)  & = &  \max_{ \Sigma \in \widehat{\mathcal{K}}}\ \  \tr [ \Sigma M] \ \ = \ \ \max_{ \Sigma \in \mathbb{H}_d } \ \ \tr [ \Sigma M]  \ \ \mbox{ such that }\  \tr [ \Sigma    ] = 1, \ \Sigma \in  \mathcal{V} , \mbox{ and } \Sigma \succcurlyeq 0,
\EEA
which satisfies $\Gamma(M) \leqslant \widehat{\Gamma}(M)$ for all $M \in \mathbb{H}_d$. 

These relaxations are often referred to as ``sum-of-squares'' (SOS) relaxations, because of the following dual interpretation.\footnote{{Note that using an outer approximation in \eq{gammaaa} leads to a relaxation \emph{per se}, while in the dual view presented here, we obtain a \emph{strengthening} due to replacing non-negative functions by the smaller set of sums-of-squares. Throughout the paper, we will use the term ``relaxation'' in all cases.}} Introducing Lagrange multipliers, $c \in \rb$ for the constraint $\tr [ \Sigma   ] = 1$, $Y \in \mathcal{V}^\perp$ for $\Sigma \in  \mathcal{V}$, and $B \succcurlyeq 0$ for $\Sigma \succcurlyeq 0$, we get, using strong duality {(which holds by Slater's condition since \eq{gammabef} has a strictly feasible point)}:
\BEA
\notag \widehat{\Gamma}(M) & = &  \sup_{  \Sigma \in \mathbb{H}_d} \ \inf_{ c \in \rb , \ Y \in \mathcal{V}^\perp, \ B \succcurlyeq 0} \ 
\tr [ \Sigma M ] + c ( 1 - \tr[ \Sigma  ] ) + \tr  [Y \Sigma] + \tr [B \Sigma] \\
 \label{eq:pd}   & = &    \inf_{ c \in \rb , \ Y \in \mathcal{V}^\perp, \ B \succcurlyeq 0} \  \ \sup_{  \Sigma \in \mathbb{H}_d} \ 
\tr [ \Sigma M ] + c ( 1 - \tr[ \Sigma  ] ) + \tr [ Y  \Sigma] + \tr [B \Sigma] \\
\label{eq:gammabef} & = &    \inf_{ c \in \rb , \ Y \in \mathcal{V}^\perp, \ B \succcurlyeq 0} c \ \ \mbox{ such that } \ M =  c \idm - Y - B \\
 \notag & = &    \inf_{ c \in \rb , \ B \succcurlyeq 0} c \ \ \mbox{ such that }\  \forall x \in \X, \  c - \varphi(x)^\ast M \varphi(x)  =   \varphi(x)^\ast B \varphi(x).
\EEA
This can be interpreted as finding the lowest upper-bound $c$ on the function $x \mapsto \varphi(x)^\ast M \varphi(x) $ by relaxing the non-negativity of $ c - \varphi(x)^\ast M \varphi(x)  $ by  the existence of $B \succcurlyeq 0$ such that  $c - \varphi(x)^\ast M \varphi(x)  =   \varphi(x)^\ast B \varphi(x)$ (which is indeed non-negative). Finally, using the eigendecomposition of $B$, $\varphi(x)^\ast B \varphi(x)$ can be written as a sum of square functions, hence the denomination. Note that it is common to add extra conic constraints to further restrict $\widehat{\mathcal{C}}^\ast$, often leading to hierarchies of relaxations (see examples below and~\cite{lasserre2010moments}), which makes the relaxations tighter and tighter. {This corresponds to adding to $\varphi$ another feature vector $\varphi^+$, and see a quadratic form in $\varphi$ as a quadratic form in $\tilde{\varphi} = { \varphi \choose \varphi^+}$, which leads to a tighter relaxation.}

{In terms of computational complexity, because Slater's condition is satisfied, interior-point methods have a polynomial number of iterations, each based on polynomial-time numerical linear algebra algorithms~\cite{nesterov1994interior}.}

\paragraph{Spectral relaxation.}{Because of our unit norm assumption for the features, we can maximize with respect to $c$ and $B$ in \eq{gammabef} and reformulate
\eq{gammabef} as the problem of minimizing $\lambda_{\max}(M+Y)$ over $Y \in \mathcal{V}^\perp$. Simply taking $Y=0$ corresponds to computing the largest eigenvalue of $M$. This corresponds to relaxing $\mathcal{K}$ to $\{ \Sigma \succcurlyeq 0, \ \tr[\Sigma]=1\}$.}

Throughout the paper, we will often use the following statements based on dual cones (using that the dual of the intersection of {closed cones is the sum of their duals when their relative interiors intersect~\cite[Corollary~16.4.2]{rockafellar2015convex}, and the assumption that $\mathcal{V}$ contains a positive definite matrix)}:
\BEA
\notag {\Gamma}(M)  \leqslant t  &  \Leftrightarrow & \Gamma( t \idm - M ) \leqslant 0 \ \  \Leftrightarrow  \ \  t\idm - M \in  {\mathcal{C}}^\ast  \\
\label{eq:A}
\widehat{\Gamma}(M)  \leqslant t  &  \Leftrightarrow &  \widehat\Gamma( t\idm - M ) \leqslant 0   \ \  \Leftrightarrow \ \   t\idm - M \in \widehat{\mathcal{C}}^\ast =  \mathbb{H}_d^+ +  \mathcal{V}^\perp \\
\notag &  \Leftrightarrow &  {\exists Y \in \mathcal{V}^\perp, \ \lambda_{\max}(M+Y) \leqslant t }.
\EEA

\subsection{Examples}
\label{sec:examples}

\paragraph{Finite set with injective embedding.} If $\X$ is finite and the Gram matrix of all features for all values of~$\X$ is invertible, then the SOS relaxation is tight. Indeed, assuming (potentially after applying an invertible linear transformation to $\varphi$) that $\varphi(x)^\ast \varphi(y) = 1_{y=x}$, $\mathcal{K}$ is the set of diagonal matrices with a diagonal belonging to the simplex.

\paragraph{Affine functions on the Euclidean unit sphere.}
We consider $\X$ the unit sphere in $\rb^{d-1}$, with $\ds \varphi(x) = \frac{1}{\sqrt{2}} {1 \choose x} \in \rb^{d}$. Then $\mathcal{V}$ is the set of matrices $\bigg( \begin{array}{ll}\! \!\alpha & x^\top\!\! \\ \!\! x & X\!\!
\end{array} \bigg)$ such that $\tr(X) = \alpha$. This is another situation where the sum-of-squares relaxation is tight~\cite{hager2001minimizing}.

\paragraph{Trigonometric polynomials on $[0,1]$.}
We consider $\X = [-1,1]$ and $\varphi(x) \in \cb^{2r+1}$, with $\varphi(x)_\omega = e^{ 2i\pi \omega x}/ \sqrt{2r+1}$ for $\omega \in \{-r,\dots,r\}$. Then $(\varphi(x)\varphi(x)^\ast)_{\omega \omega'} = e^{2 i\pi (\omega - \omega') x} / (2r+1)$, and thus $\mathcal{V}$ is the set of Hermitian  Toeplitz matrices. It turns out that the sum-of-squares relaxation is tight, see \cite[Theorem 1.2.1]{szego1975orthogonal} and \cite{dumitrescu2007positive}.

\paragraph{Trigonometric polynomials on $[0,1]^n$.} We consider $\X = [-1,1]^n$ and $\varphi(x)_\omega = e^{2 i\pi \omega^\top x} / (2r+1)^{n/2} \in \cb$ for $\omega$ in a certain set $\Omega \subset \zb^n$, typically $\Omega = \{ \omega \in \zb^n, \ \| \omega\|_\infty \leqslant r\}$. We then have 
$(\varphi(x)\varphi(x)^\ast)_{\omega \omega'} = e^{ 2 i\pi (\omega - \omega')^\top x}  / (2r+1)^{n}  $, which depends only on $\omega - \omega'$, which defines a set of linear constraints defining $\mathcal{V}$. The relaxation is then not tight, but by embedding $\Omega$ in a larger set, we can make the relaxation as tight as desired~(see~\cite{dumitrescu2007positive}), {while for $n=2$, it will be tight after a certain (unknown) degree~\cite[Corollary 3.4]{scheiderer2006sums}.}

\paragraph{Polynomials on $[-1,1]$.} {In order to tackle polynomials, we could simply consider $\varphi(x)$ composed of monomials, but this would not lead to a normalized feature map. Rather, following \cite[Section 2.2]{bach2022exponential}, given a classical polynomial $P$, we consider the trigonometric polynomial $x\mapsto f(x) = P(\cos 2\pi x)$, which we can represent with a normalized feature map. This extends to polynomials on $[-1,1]^n$.}

\paragraph{Polynomials on the Euclidean hypersphere.} {With $\X = \{x \in \rb^{n+1}, x^\top x = 1\}$, we can consider all harmonic polynomials, as described by~\cite{fang2021sum}. By only considering functions depending on the first $n$ variables, this allows to consider $\X$ the Euclidean unit ball.}

\paragraph{Boolean hypercube.}
We consider 
  $\X = \{-1,1\}^n $  with feature vectors composed of Boolean Fourier components of increasing orders~\cite{o2014analysis}. This corresponds to features
   $\varphi_A(x) = \prod_{i \in A} x_i \in \{-1,1\}$, where $A$ is a subset of $\{1,\dots,n\}$.   Moreover, given two sets $A$ and $B$, we have $\varphi_A(x) \varphi_B(x) = \varphi_{A \Delta B}(x)$, where $A \Delta B$ is the symmetric difference between $A$ and $B$.
   
If we consider a set $\mathcal{A}$ of subsets of $\{1,\dots,n\}$, then, the element indexed $(A,B)$ of $\varphi(x) \varphi(x)^\ast$ only depends on the symmetric difference $A \Delta B = ( A \backslash B) \cup ( B \backslash A)$, and this leads to a set of linear constraints defining~$\mathcal{V}$. The relaxation is not tight in general, but if we see our moment matrix as a submatrix obtained from a sufficiently larger set of subsets, then we obtain a tight formulation 
(see~\cite{lasserre2001explicit,laurent2003comparison,slot2022sum} and references therein).

\section{Exact lower bounds based on moments}
\label{sec:exact}
 
We consider the optimal lower bound on $D(p\|q)$ given the integrals $  \Sigma_p = \int_{\X} \varphi(x)\varphi(x)^\ast dp(x)$ and $ \Sigma_q = \int_{\X} \varphi(x)\varphi(x)^\ast dq(x)$,\footnote{{Note that $\Sigma_p$ and $\Sigma_q$ do depend on $\varphi$, but we omit the dependence in the notation.}} that is, given two Hermitian matrices $A$ and $B$: 
\BEQ
\label{eq:DOPT}
D^{\rm OPT} (A\|B) = \inf_{p, q \in \mathcal{M}_+(\X)}\  D(p\|q) \ \mbox{ such that } \ \Sigma_p = A  \mbox{ and } \Sigma_q = B.
\EEQ
{Note that we do not assume that $p$ and $q$ integrates to one.}
By construction, $D^{\rm OPT}( \Sigma_p \| \Sigma_q) \leqslant D(p\|q)$. Moreover, we have some immediate properties for the function of $(A,B)$ defined in \eq{DOPT}, which preserves similar properties of $D(p\|q)$. For other potential properties such as used within multivariate probabilistic modeling, see~\cite{bach2022information}:
\BIT
\item If $A$ or $B$ is not in $\mathcal{C}$ (the closure of the convex hull of all $\varphi(x) \varphi(x)^\ast$), then 
{the optimization problem in \eq{DOPT} is infeasible as no $(p,q)$ can be found to satisfy $\Sigma_p=A$ and $\Sigma_q=B$, and, following the standard convention in convex analysis~\cite{rockafellar2015convex}, we set the value of $D^{\rm OPT} (A\|B)$ to infinity.}

\item If $\tr [ A  ] = \tr [B ] = 1$, since $\varphi$ is normalized to unit norm, then the optimal measures $p$ and $q$ in \eq{DOPT} are probability measures.
\item The function $(A,B) \mapsto D^{\rm OPT} (A\|B) $ is \emph{jointly} convex in $A$ and $B$, as the optimal value of a jointly convex problem in $A,B,p,q$.
\item If $\varphi$ is replaced by $T \varphi$ for an injective linear map $T$ {(which changes also $\Sigma_p$ and $\Sigma_q$)}, the quantity $D^{\rm OPT} (\Sigma_q\|\Sigma_q)$  is unchanged.\footnote{{In this section, we do not need to impose that $\| T \varphi(x)\| = 1$ for all $x \in \X$, but we will in subsequent sections.}}

\item If $\varphi$ is replaced by $T \varphi$ for a (potentially non-injective) linear map $T$, the quantity $D^{\rm OPT} (\Sigma_q\|\Sigma_q)$ is reduced. Therefore, to have tighter lower-bounds $D^{\rm OPT} (\Sigma_q\|\Sigma_q)$ on $D(p\|q)$, we need to use high-dimensional features. In other words, for all $p,q$, the approximation is typically tight, that is, $D(p\|q)$ close to $ D^{\rm OPT}(\Sigma_p\|\Sigma_q)$ only if the feature $\varphi: \X \to \cb^d$ is rich enough. For approximation capabilities when the feature size grows to infinity, and the use of positive definite kernel methods, see~\cite{bach2022information}. In this paper, all feature vectors will have a fixed dimension.

 \EIT

\paragraph{Variational representation.}
We have, using the representation of the $f$-divergence $D(p\|q)$ from \eq{f-var}, and strong convex duality for an infinite-dimensional optimization problem with linear constraints~\cite[Section~8.6]{luenberger1997optimization}, {with Lagrange multipliers $M,N \in \mathbb{H}_d$, for the finite-dimensional equality constraints $A = \int_\X \varphi(x)\varphi(x)^\ast dp(x)$
and $B = \int_\X \varphi(x)\varphi(x)^\ast dq(x)$}:  
\BEA
\nonumber &&D^{\rm OPT} (A\|B)  \\
\nonumber& = & \inf_{ p,q \in   \mathcal{M}(\X) } \ \ \sup_{M,N \in \H_d , \ v,w : \X \to \rb }
\tr \Big[ M \Big( A - \int_\X     \varphi(x)  \varphi(x)^\ast  dp(x) \Big) \Big] + \tr \Big[ N \Big( B - \int_\X     \varphi(x)  \varphi(x)^\ast  dq(x) \Big) \Big] \\
\nonumber&&  \hspace*{2cm} + \int_\X v(x) dp(x)  + \int_\X w(x) dq(x)   \mbox{ such that } \ \forall x \in \X, \forall r \geqslant 0, \  r v(x)  + w(x) \leqslant f(r)\\
\nonumber& = & \sup_{M,N \in \H_d , \ v,w : \X \to \rb }\ \ 
\inf_{ p,q \in   \mathcal{M}(\X) } 
\tr [ M A] + \tr [ NB]  + \int_\X   \big( v(x) - \varphi(x)^\ast M \varphi(x)\big)  dp(x)
\\
\nonumber&&  \hspace*{2cm} +  \int_\X \big( w(x)  - \varphi(x)^\ast N \varphi(x)\big) dq(x)   \mbox{ such that } \ \forall x \in \X, \forall r \geqslant 0, \  r v(x)  + w(x) \leqslant f(r)\\
\nonumber & & \hspace*{12cm} {\mbox{by swapping $\inf$ and $\sup$,}}
\\
\label{eq:opty} & = & \sup_{M,N\in \H_d }\ 
\tr [ M A ] + \tr [ NB]  \ \mbox{ such that } \forall x \in \X, \forall r \geqslant 0, \  r   \varphi(x)^\ast M \varphi(x)    +   \varphi(x)^\ast N \varphi(x)  \leqslant  f(r)\\
\nonumber & & \hspace*{1.2cm} {\mbox{since taking the infimum with respect to } p \mbox{ and } q \mbox{ leads to explicit expressions for } v \mbox{ and } w,}
\\
\label{eq:DSOS} & = & \sup_{M,N \in \mathbb{H}_d}\
\tr [ M A ] + \tr [ NB]  \ \mbox{ such that } \ \forall r \geqslant 0, \ \Gamma(rM+N) \leqslant  f(r), {\mbox{ by definition of } \Gamma}.\EEA
{Note that above, we can restrict $M$ and $N$ to be in $\mathcal{V}$ by orthogonal projection on $\mathcal{V}$, as any component along $\mathcal{V}^\perp$ does not change the optimization problem.}
The representation above shows that being able to compute $D^{\rm OPT}$ requires the computability of $\Gamma$ defined in \eq{gamma}, that is, maximizing quadratic forms in $\varphi(x)$, which is exactly what SOS methods presented in \mysec{sos} are tailored to approximate, and that will be used in \mysec{sos-relax} below.

\paragraph{Algorithms to compute $D^{\rm OPT} (A\|B) $.}

This tightest lower bound can only be {computed precisely}, if we can compute $\Gamma$ arbitrarily precisely. This is typically only easily possible without brute force enumeration with sum-of-squares relaxations which are asymptotically tight (thus using hierarchies in dimensions larger than one). In our experiments where we compare all bounds, we consider the case of uni-dimensional trigonometric polynomials, for which our simplest relaxation is already tight.
Computable lower bounds are considered in \mysec{sos-relax} (based on SOS relaxations) and \mysec{qt-relax} (based on quantum information divergences).

\paragraph{Alternative derivation.} {While we gave a definition of $D^{\rm OPT} ( A \| B ) $ in terms of smallest possible $D(p\|q)$ given $\Sigma_p = A$ and $\Sigma_q = B$, the formulation in \eq{opty} shows that it is equivalent to taking the variational formulation of $f$-divergences in \eq{f-var}, and only allowing quadratic forms in $\varphi$ for the functions $w$ and $v$, that is, $w(x) = \varphi(x)^\ast M \varphi(x)$ and $v(x) = \varphi(x)^\ast N \varphi(x)$.}

\paragraph{Relaxations.}
{What makes the computation of $D^{\rm OPT} (A\|B) $ difficult is the need to deal with the constraints $\Gamma(rM+N)  \leqslant f(r) $ in \eq{DSOS}.
In the next two sections, we will explore two successive relaxations:}
\BIT
\item {The first one in \mysec{sos-relax} corresponds to using the constraints
$\widehat{\Gamma}(rM+N)  \leqslant f(r) $, which is thus using the SOS relaxation for the optimization problem.
This is equivalent to the existence of $Y^{(r)} \in \mathcal{V}^\perp$ (one for each $r$) such that $rM + N + Y^{(r)} \preccurlyeq f(r) \idm$.
Equivalently, this is using $\widehat{\mathcal{C}}$ instead of $\mathcal{C}$, and will lead to the lower-bound $D^{\rm SOS} (A\|B) $.}
\item {
The second one in \mysec{qt-relax} adds a further relaxation by having a unique $Y$ independent of $r$, in a specific form that leads to a direct link with spectral quantum divergence from \mysec{qt-review}. This will lead to the lower-bound $D^{\rm QT} (A\|B) $.
}
\EIT

\section{Relaxed $f$-divergence based on SOS}
\label{sec:sos-relax}

We consider replacing $\Gamma$    in  the optimal relaxation $D^{\rm OPT}(A\|B)$ in \eq{DSOS} by its approximation $\widehat{\Gamma}$ based on sums-of-squares, as defined in \mysec{sos}. This leads to, using  that   $\widehat{\mathcal{C}}  = \mathbb{H}_d^+  \cap  \mathcal{V}$, and thus $\widehat{\mathcal{C}} ^\ast = \mathbb{H}_d^+ +  \mathcal{V}^\perp$, from \eq{A}: 
\BEA
\label{eq:ZZ} D^{\rm SOS} ( A \| B ) 
& = & \sup_{M,N \in \mathbb{H}_d} \ \ \tr [ AM ] + \tr [ BN] \ \mbox{ such that } \ \forall r \geqslant 0, \ \widehat{\Gamma}(rM+N) \leqslant f(r)
\\
\notag & = & \sup_{M,N \in  \mathbb{H}_d} \ \ \tr [ AM ] + \tr [ BN]\  \mbox{ such that }\  \forall r \geqslant 0, \ f(r)\idm  -  rM-N \in \widehat{\mathcal{C}}^\ast = \mathbb{H}_d^+ +  \mathcal{V}^\perp .
\EEA
Since $\Gamma \leqslant \widehat{\Gamma}$, we have by construction $D^{\rm SOS} \leqslant D^{\rm OPT}$.  {Most of the properties of $D^{\rm OPT}$ are preserved, such as convexity, invariance by invertible transforms. In terms of domain,} it is now finite if only if $A,B \in \widehat{\mathcal{C}} = \H_d^+ \cap \mathcal{V}$ (rather than $\mathcal{C}$).

\paragraph{Variational unconstrained formulation.}  {In order to derive an algorithm to estimate $D^{\rm SOS} ( A \| B ) $, we will use the same unconstrained variational formulation as \eq{AAA},\footnote{{A similar formulation can be derived for $D^{\rm OPT}$  using $\Gamma$ instead of $\widehat{\Gamma}$.}} and introduce the function:
\BEAS
 \widehat{G} (M,N) & = &  \sup_{\Sigma \in \widehat{\mathcal{K}}} F\big(\tr[M\Sigma],\tr[N\Sigma]\big)
 =\sup_{\Sigma \in \widehat{\mathcal{K}}}  \sup_{r \geqslant 0} 
  \frac{ r \tr[M\Sigma] + \tr[N\Sigma] - f(r)}{r+1} \\
 & = & \sup_{r \geqslant 0} \frac{\widehat{\Gamma}( rM +  N )  -  f(r) }{r+1}
=  \sup_{s \in [0,1]} \widehat{\Gamma}( sM + (1-s)N) - (1-s) f\big( \frac{s}{1-s}\big),
 \EEAS
 by the change of variable $s = r/(r+1) \Leftrightarrow r = s/(1-s)$.
We can therefore use the same reasoning as the one leading to \eq{AD}, now based on $ \widehat{G} (M-u\idm,N-u
\idm)= \widehat{G} (M,N) -u$, and obtain from \eq{ZZ} the unconstrained variational formulation:
\BEA
\notag D^{\rm SOS} ( A \| B ) 
 & = & \sup_{M,N \in \mathbb{H}_d} \ \ \tr [ AM ] + \tr [ BN] 
\ \mbox{ such that }   \ \widehat{G} (M,N) \  \leqslant 0 \\
\label{eq:GMN} & = & \sup_{M,N \in \mathbb{H}_d} \ \ \tr [ AM ] + \tr [ BN] 
- ( \tr[A]+\tr[B])   \widehat{G} (M,N) . \EEA}

\paragraph{Estimation algorithm based on Kelley's method.} {
Given the unconstrained formulation above, we are faced with the maximization of a convex function over a vector space. Assuming that subgradients of~$\widehat{G}$ can be computed, we can use Kelley's method~\cite{kelley1960cutting}, 
which is constructing a sequence of piecewise-affine lower-bounds on $\widehat{G}(M,N)$ obtained from subgradients and use them to find the next candidate maximizer for $M,N \in \mathbb{H}_d$. Note that we can restrict the search for $M-M_0$ and $N-N_0$ in $\mathcal{V}$  (for any well-chosen $M_0$, $N_0$ typically from the computationally more efficient relaxations derived in \mysec{qt-relax})  since any component along $\mathcal{V}^\perp$ has no impact on the objective function. Since we do not know a priori an upper bound on $\| N - N_0\|$ and $\| M - M_0\|$, we restrict the minimization to bounded sets, and increase the bound if the boundedness constraints are active.  }

\paragraph{Computing subgradients.} {
 We need to find maximizers defining the convex function  $\widehat{G}(M,N)$ above, that is, finding maximizer $s$ and $\Sigma$  in
 $\sup_{\Sigma \in \widehat{\mathcal{K}}}  \sup_{s \in [0,1] } 
   \tr[ \Sigma(sM + (1-s)N)] -  (1-s) f\big( \frac{s}{1-s}\big)$. Given $s$, this is equivalent to solving one SOS maximization problem. We use the Chambolle-Pock algorithm~\cite{chambolle2011first} for the primal-dual formulation in \eq{pd}, with a fixed maximal number of iterations. Since we end up solving many such problems for values of $s$ which are close, we use warm starts and simple predictor-corrector steps so that the maximal number of iterations is rarely achieved.
  
  To find an approximate joint maximizer $(s,\Sigma$), we first take a fine grid $s \in [0,1]$, and compute the value of $\widehat{\Gamma}( sM + (1-s)N) - (1-s) f\big( \frac{s}{1-s}\big)$
  for each element of this grid. From the largest elements, we launch alternating optimization (alternating optimizing over $s$ and over $\Sigma$), which is converging quickly. The number of steps can be controlled if needed, but we simply take a grid of step less than $10^{-3}$.}

\paragraph{Using hierarchies.} {In order to get tighter approximations to $D^{\rm OPT}$, like in classical SOS optimization, we can embed the feature $\varphi$ into a larger feature vector $\tilde{\varphi} = { \varphi \choose \varphi^+}$, where $\varphi^+$ is an additional feature map. This defines the expression $D^{\rm SOS}_{\tilde{\varphi}}(\tilde{A}\| \tilde{B})$, for moment matrices $\tilde{A}$ and $\tilde{B}$, where we add the dependence on the feature map to makes the difference explicit. We can 
then the consider a tighter bound by minimizing $D^{\rm SOS}_{\tilde{\varphi}}(\tilde{A}\| \tilde{B})$ with respect to $\tilde{A}$ and $\tilde{B}$ that have their upper-left blocks equal to $A$ and $B$.}

In this paper, we focus on the computation of upper-bounds of the partition function. The study of the approximation capabilities, when the feature vector grows, is left for future work.

\section{Relaxed $f$-divergence based on quantum information theory}

\label{sec:qt-relax}

{In the previous section, we relaxed $\Gamma$ into $\widehat{\Gamma}$, which corresponds to using SOS relaxations for the optimization of quadratic forms. This was equivalent to replacing $\mathcal{C}$ by $\widehat{\mathcal{C}}$. This can be further relaxed into the spectral relaxation where we replace
$\widehat{\Gamma}$ by the spectral relaxation $\lambda_{\max}$ (which is larger), which is equivalent to replacing $\widehat{\mathcal{C}}$ by the PSD cone $\mathbb{H}_d^+$ (which is larger).}

{Starting from \eq{ZZ}, this thus leads to:
\BEA
\label{eq:QTmaxx} &  &  \sup_{M,N \in \H_d } \ 
 \tr [ MA ] + \tr [ NB] \  \mbox{ such that }  \  \forall r \geqslant 0, \   r M  + N \preccurlyeq f(r) \idm.  
 \EEA
The following lemma, taken from~\cite[Section 9.1]{matsumoto2015new} and with a proof shown in Appendix~\ref{app:lemma}, shows that we obtain exactly the ``maximal'' quantum information divergence $\tilde{D}^{\rm QT}_{\max}(A\|B)$ defined in \mysec{qt-review}.}

\begin{lemma}\cite[Section 9.1]{matsumoto2015new}
\label{lemma:ABL}
Assume $A, B \succcurlyeq 0$  and $f$ is operator-convex. The maximal value of \eq{QTmaxx} is equal to 
$\tr \big[ B f \big( B^{-1/2}A B^{-1/2} \big) \big]$ (see proof for minimizer).
\end{lemma}

{The simplest spectral relaxation thus satisfies $ \tilde{D}^{\rm QT}_{\max}(A\|B) \leqslant   {D}^{\rm SOS} (A\|B)$. Note that this is already an improvement on \cite{bach2022information}, which considers the larger ``standard'' quantum divergence.}

{In the relaxation $ \tilde{D}^{\rm QT}_{\max}$, we keep the joint convexity, but we lose the partial invariance by invertible linear transform that both $D^{\rm OPT}$ and $D^{\rm SOS}$ had. This can be remedied by defining $ {D}^{\rm QT} $ as the largest possible value of 
$\tilde{D}^{\rm QT}_{\max}$ once these invariances are taken into account. This link will be explored in \mysec{linkQT},  and for now we define $D^{\rm QT}$ as follows:
 \BEQ
 \label{eq:VV}
 \!D^{\rm QT}(A\|B)  =
 \sup_{M,N,V \in \H_d} \tr [MA] + \tr [ NB] \mbox{ such that } \forall r \geqslant 0, r M + N \preccurlyeq f(r) \textcolor{red}{V}, \ V \succcurlyeq 0, \ V  - \idm \in \mathcal{V}^\perp,
 \EEQ
 where the only difference with \eq{QTmaxx} is the introduction of the matrix $V$ which can be different from~$\idm$. Comparing to \eq{ZZ}, we get that it is between  $\tilde{D}^{\rm QT}_{\max}(A\|B)$ and $D^{\rm SOS}(A\|B)  $, because the constraint $r M + N \preccurlyeq f(r) V$ in \eq{VV} implies $f(r)\idm  -  rM-N \in \mathbb{H}_d^+ +  \mathcal{V}^\perp$ (and we thus get a feasible point for \eq{ZZ}).}
 
 {It turns out that we can also optimize in closed form with respect to $M$, $N$ leading to a similar expression for 
 $D^{\rm QT}(A\|B)$ in \eq{VV}. The following lemma is proved in Appendix~\ref{app:lemma}, and is a slight modification of Lemma~\ref{lemma:ABL}.}
 
 \begin{lemma}
 \label{lemma:ABLV}
 Assume $A, B \succcurlyeq 0$  and $f$ is operator-convex. The maximal value of \eq{VV} is equal to:
 \BEQ
\label{eq:QTdef} D^{\rm QT}(A\| B)
=
 \sup_{V \succcurlyeq 0, \ V  - \idm \in \mathcal{V}^\perp}\  \tr \big[  V B^{1/2}  f \big( B^{-1/2}A B^{-1/2} \big) B^{1/2} \big].
\EEQ
 \end{lemma}
 We can thus obtain the value of $D^{\rm QT}(A\| B)$ by a convex optimization problem that does not involve any spectral function, is finite-dimensional, and can be solved using any semi-definite programming solver.

\paragraph{Properties of $D^{\rm QT}(A\| B)$.} We start by deriving an alternative formulation.
Defining the matrix $Q = B^{1/2}  f \big( B^{-1/2}A B^{-1/2} \big) B^{1/2} \succcurlyeq 0 $, we get:
\BEAS
  D^{\rm QT}(A\|B) 
& =  & \max_{V}\  \tr [QV ]  \mbox{ such that }   V \succcurlyeq 0 \mbox{ and } \forall x \in \X, \   \varphi(x)^\ast V \varphi(x)  = 1 \\
& = & \max_{V} \ \tr [QV ]  \mbox{ such that }   V \succcurlyeq 0 \mbox{ and } V - \idm \in \mathcal{V}^\perp \\
& = & \min_{\Sigma \in \mathcal{V}} \ \tr [\Sigma  ]  \mbox{ such that }   \Sigma \succcurlyeq Q, \mbox{ by Lagrange duality,} \\
& = &  \min_{\Sigma \in \mathcal{V}} \ \tr [\Sigma  ]  \mbox{ such that }   \Sigma \succcurlyeq  B^{1/2}  f \big( B^{-1/2}A B^{-1/2} \big) B^{1/2}.
 \EEAS

 Note that like $D^{\rm OPT}$ and $D^{\rm SOS}$, $D^{\rm QT}$ is jointly convex, and  $D^{\rm QT}(\Sigma_p\|\Sigma_q)$ is invariant by invertible linear transform of $\varphi$ (which would not be the case without optimizing with respect to $V$) (see \mysec{linkQT}). Moreover, $D^{\rm QT}(A\|B)$ is finite only if $A$ and $B$ are positive semi-definite (as opposed to be also in $\mathcal{V}$ for $D^{\rm SOS}(A\|B)$).

 In terms of algorithms to approximate $D^{\rm QT}(A\| B) $, we can use interior-point methods to solve \eq{QTdef} as this is not a computational bottleneck.

\paragraph{Unconstrained formulation.}{In order to compute the lower-bound on the $f$-divergence $ D^{\rm QT}$ the formulations above are the most appropriate. However, when dealing with variational inference in \mysec{qt-part}, we will the need the following unconstrained formulation, akin to \eq{GMN}, and obtained by replacing $\widehat{\Gamma}$ by $\lambda_{\max}$:
  \BEA
  \notag
 D^{\rm QT}(A\|B)  & = & 
 \sup_{M,N,V \in \H_d} \tr [MA] + \tr [ NB] 
 -( \tr[A] + \tr[B] ) \sup_{ r \geqslant 0} \frac{ \lambda_{\max}(r M+ N - f(r) V)}{r+1}
  \\
\label{eq:GMNQT}  & & \hspace*{7cm} \mbox{ such that } \ V \succcurlyeq 0, \ V  - \idm \in \mathcal{V}^\perp.
 \EEA
  }

 \subsection{Link with quantum information theory and metric learning}
 \label{sec:linkQT}
 Given a fixed feature map $\varphi: \X \to \cb^d$, we consider an invertible matrix $T \in \cb^{d \times d} $, such that $V = T^\ast T $ is such that  $\varphi(x)^\ast V \varphi(x) = 1$ for all $x \in \X$. {This corresponds to \emph{invertible} matrices $V$ such that $V \succcurlyeq 0$ and $V-\idm \in \mathcal{V}^\perp$.}
 Writing $\tilde{A} = T AT^\ast$, and $\tilde{B} = T BT^\ast$, we get, using the \emph{maximal divergence} defined in \mysec{qt-review}:
 $$
 \tilde{D}^{\rm QT}_{\max}(TAT^\ast\|TBT^\ast) = \tilde{D}^{\rm QT}_{\max}(\tilde{A}\|\tilde{B}) = \tr ( \tilde{B}^{1/2} f(\tilde{B}^{-1/2} \tilde{A} \tilde{B}^{-1/2})\tilde{B}^{1/2}).$$
 Since $\tilde{B}^{1/2}$ and $T B^{1/2} $ are two square roots of $\tilde{B}$, there exists a  unitary matrix $R$ such that $\tilde{B}^{1/2} = T B^{1/2} R$.
 We then get
$\tilde{B}^{-1/2} \tilde{A} \tilde{B}^{-1/2} = R^\ast B^{-1/2} A B^{-1/2} R$, leading to 
$f(\tilde{B}^{-1/2} \tilde{A} \tilde{B}^{-1/2}) = R^\ast f(B^{-1/2} A B^{-1/2}) R$, which in turn leads to
\BEAS
\tilde{D}^{\rm QT}_{\max}(\tilde{A}\|\tilde{B}) & = & 
\tr ( \tilde{B}^{1/2} f(\tilde{B}^{-1/2} \tilde{A} \tilde{B}^{-1/2})\tilde{B}^{1/2}) = 
  \tr \big[ T B^{1/2} f( B^{-1/2} A B^{-1/2}) B^{1/2} T^\ast  \big]\\
  &=&    \tr \big[ T^\ast T B^{1/2}  f( B^{-1/2} A B^{-1/2}) B^{1/2}  \big]
=  \tr \big[ V  B^{1/2}  f( B^{-1/2} A B^{-1/2}) B^{1/2}  \big] ,
\EEAS
which is exactly the objective function maximized to define $D^{\rm QT}(A\|B) $ in \eq{QTdef}. {Thus, by optimizing over all matrices $T$, and thus with respect to all matrices $V$, we have:
$$
D^{\rm QT}(A\|B) = \sup_{T \in \cb^{d\times d}, \  \forall x \in \X , \| T \varphi(x)\|=1}  \tilde{D}^{\rm QT}_{\max}(TAT^\ast\|TBT^\ast) .
$$}

As observed in~\cite{bach2022information} for $\tilde{D}^{QT}_{\rm standard}$,  we have  $\tilde{D}^{\rm QT}_{\max} (TAT^\ast\|TBT^\ast)  \leqslant D(p\|q)$ for any $p,q \in \mathcal{M}_+(\X)$ such that $\Sigma_p = A$ and $\Sigma_q =B$, as soon as $T^\ast T \in \idm + \mathcal{V}^\perp$. Our new relaxation is thus equivalent to estimating the best feature vector in a linear model defined by $\varphi$. A simple consequence is that, while $\tilde{D}^{\rm QT}_{\max}$ is not invariant by invertible linear transforms, $D^{\rm QT}$ is (just like $D^{\rm OPT}$ and $D^{\rm SOS}$). Note finally, that the use of $\tilde{D}^{QT}_{\rm standard}$ instead of $\tilde{D}^{QT}_{\rm max}$, as done in~\cite{bach2022information} for the particular case of the KL divergence, leads to a weaker relaxation and a more complex optimization problem in $V$ (concave maximization instead of linear maximization).

 \section{Variational inference with $f$-divergences}
 \label{sec:varinf}
 
 {Now that we have explored convex lower-bounds for the $f$-divergence, we can explore the Fenchel-dual equivalent, that is, convex upper-bounds on $f$-partition functions, and the natural link with variational inference. We start with a review of probabilistic concepts  without any feature map in  \mysec{fpart} and then with a feature map in \mysec{varfeat}, before exploring relaxations in subsequent sections.}
 
 \subsection{$f$-partition function}
 \label{sec:fpart}

  Given a function $h: \X \to \rb$, and $q $ a fixed positive measure not necessarily summing to one (that is, in $\mathcal{M}_+(\X)$), we can define the ``$f$-partition function'' as the Fenchel dual with respect to {a probability measure}~$p$ of $D(p\|q)$, that is:
   \BEQ
   \label{eq:donsker}
   c_q(h)   = \sup_{ p \in \mathcal{P}(\X)} \ \int_\X h(x) dp(x) - D(p\|q).
   \EEQ
   Using the variational formulation in \eq{f-var}, {we can optimize with respect to $w$ to obtain
   \BEAS
  c_q(h)   & = & \sup_{p   \in \mathcal{M}(\X) } \inf_{v: \X \to \rb} \ \int_\X h(x) dp(x)  - \int_\X v(x) dp(x) +  \int_\X f^\ast(v(x)) dq(x) \mbox{ such that } \int_\X dp(x) = 1,
  \EEAS
  where it is sufficient to consider $p   \in \mathcal{M}(\X)$ rather than $p   \in \mathcal{M}_+(\X)$ because the infimum is only finite for non-negative measures. We can then introduce a Lagrange multiplier $\rho$ for the single equality constraint $ \int_\X dp(x) = 1$, and using strong duality to swap infimum and supremum~\cite[Section~8.6]{luenberger1997optimization},} we get (see~\cite{picard2022change} for similar derivations in the context of PAC-Bayes analysis):
  \BEAS
  c_q(h)    & = & \sup_{p  \in \mathcal{M}(\X)} \ \inf_{\rho \in \rb, \  v: \X \to \rb}\  \int_\X h(x) dp(x)  -   \int_\X v(x) dp(x) +  \int_\X f^\ast(v(x)) dq(x)  - \rho \Big( \int_\X dp(x) - 1 \Big) \\
   & = & {\inf_{\rho \in \rb, \  v: \X \to \rb} \ \ \sup_{p  \in \mathcal{M}(\X)} \  \int_\X h(x) dp(x)  -   \int_\X v(x) dp(x) +  \int_\X f^\ast(v(x)) dq(x)  - \rho \Big( \int_\X dp(x) - 1 \Big).} 
   \EEAS
   {We can then optimize in closed form with respect to $p$, which leads to $v =  h - \rho$, and then with respect to~$v$, which leads to
   \BEA
\label{eq:cqhi}     c_q(h)   
  & = & \inf_{\rho \in \rb }  \ \ \rho  +\int_\X  f^\ast(h(x)-\rho) dq(x)  .
  \EEA}
The optimality condition for $\rho$ is that $  \int_\X (f^\ast)'(h(x)-\rho) dq(x) = 1$. Moreover, the set of functions $h: \X \to \rb$ such that $c_q(h)$ is finite is a convex set.
 
  This means that we can define a probability distribution with density $ (f^\ast)'(h(x)-\rho)$ with respect to $q$, which we denote $p(x|h)$.  For $f(t) = t \log t - t + 1$, where $(f^\ast)'(u) = e^u$, we recover classical exponential families (see~\cite{wainwright2008graphical,murphy2012machine} and references therein), and $c_q(h)=
    \log \big( \int_\X e^{h(x)} dq(x) \big) + 1 - \int_\X q(x)$, which is the traditional log-partition function, and \eq{donsker} is often referred to as the Donsker-Varadhan inequality~\cite{donsker1976asymptotic}.
 
\paragraph{Variational formulation.} We can use the variational formulation in \eq{f-var}, but {without first optimizing~$w$ out},  still using strong duality, {and adding a Lagrange multiplier $\rho$ for the constraint $\int_\X dp(x)=1$}:
    \BEA
 \nonumber c_q(h) & = &  \sup_{ p \in \mathcal{P}(\X)} \ \int_\X h(x) dp(x) - D(p\|q) \\
 \nonumber & = & \sup_{p  \in   \mathcal{M}(\X)} \inf_{v,w: \X \to \rb} \int_\X h(x) dp(x)  - \int_\X v(x) dp(x) -  \int_\X w(x)  dq(x) \mbox{ such that } \int_\X dp(x) = 1 \\
\nonumber  & & \hspace*{5cm} \mbox{ such that } \forall x \in \X, \forall r \geqslant 0, \  r v(x)  + w(x) \leqslant f(r)\\
\nonumber  & = & \sup_{p\in   \mathcal{M}(\X)} \ \inf_{\rho, \ v,w: \X \to \rb} \ \int_\X h(x) dp(x)  - \int_\X v(x) dp(x) -  \int_\X w(x)  dq(x)  - \rho \bigg(  \int_\X dp(x) -1 \bigg) \\
\nonumber  & & \hspace*{5cm} \mbox{ such that } \forall x \in \X, \forall r \geqslant 0, \  r v(x)  + w(x) \leqslant f(r)\\
\label{eq:www}   & = & \inf_{\rho, \ w: \X \to \rb}    \ \rho   -  \int_\X w(x)  dq(x)  \  \mbox{ such that } \forall x \in \X, \forall r \geqslant 0, \  r h(x)  + w(x) \leqslant f(r) + \rho r,
\EEA
{since the optimization with respect to $p \in \mathcal{M}(\X)$ leads to $v=h$. {Like in earlier variational formulations, we can replace the constrained formulation by an unconstrained one using the function $F$ defined in \eq{defF} in \mysec{varf}.}
By replacing $w$ by $w-\rho$, and optimizing with respect to $\rho$ in \eq{www}, we get:}
\BEA
 \nonumber c_q(h)& = & {\inf_{ w: \X \to \rb}    \Big(1 + \int_\X dq(x) \Big) \sup_{x \in \X} \sup_{r \geqslant 0} \frac{ r h(x)  + w(x) - f(r)}{r+1}  -  \int_\X w(x)  dq(x)}  \\
 & = &  {\inf_{ w: \X \to \rb}     \Big(1 + \int_\X dq(x) \Big) \sup_{x \in \X} F(h(x),w(x))  -  \int_\X w(x)  dq(x)}  ,
     \EEA
{where the optimal value of $\rho$ is obtained as $\rho = 
\sup_{x \in \X} \sup_{r \geqslant 0} \frac{ r h(x)  + w(x) - f(r)}{r+1} =  \sup_{x \in \X} F(h(x),w(x)) $. In variational inference, we will need to obtain the probability distribution $p$ in \eq{donsker} from $w$. Here, it simply has density  $ (f^\ast)'(h(x)-\rho)$ with respect to $q$.}

\subsection{Variational inference with feature maps}
\label{sec:varfeat}
{In the previous section, we have considered probability densities and partition functions for potentials that were allowed to take any functional form in $x \in \X$. We now specialize to potentials that are quadratic forms in $\varphi(x)$.}

In this section, we thus extend the notion of exponential families, which is classical for $f(t) = t \log t -t + 1$, to all $f$-divergences. These are also called ``$q$-exponential families'' for   $\alpha$-divergences~\cite{amari2011geometry}.

\paragraph{$f$-family of probability distributions.}
Following \mysec{fpart}, given the matrix feature map $x \mapsto \varphi(x) \varphi(x)^\ast \in \mathbb{H}_d$, we define the distribution $p(\cdot|H)$ with density with respect to $q$ of the form  $ (f^\ast)'\big(   \varphi(x)^\ast H  \varphi(x)    - \rho\big)$ for a certain Hermitian matrix $H \in \mathbb{H}_d$, and with the normalizing constant $\rho = \rho(H) \in \rb$ that makes the density sum to one.
We can then define 
\BEQ
\label{eq:cH}
C_q(H) = \sup_{ p\in \mathcal{P}(\mathcal{X})} \ \int_\X   \varphi(x)^\ast H \varphi(x)   dp(x) - D(p\|q)
= c_q \big(   \varphi(\cdot)^\ast H \varphi(\cdot)  \big),
\EEQ
with the optimal probability distribution $p$ {(which is unique since we have assumed that $f$ is strictly convex)} exactly being the {density $p(\cdot|H)$ defined} above. The set of $H \in \mathbb{H}_d$ such that $C_q(H)$ is finite is convex. 

From the representation {in \eq{cH}} {as a maximum of affine functions}, we obtain that
the gradient $C_q'(H)$ is equal to $  \int_{\X} p(x|H) \varphi(x)\varphi(x)^\ast dq(x)$ as $p(\cdot|H)$ is the maximizer in \eq{cH}, that is, $C_q'(H)$ is exactly the expectation of $\varphi(x)\varphi(x)^\ast$ under $p(\cdot|H)$. Thus, a classical task in variational inference is to compute $C_q'(H)$~\cite{wainwright2008graphical}. {For example, for the traditional Ising model, where $\X = \{-1,1\}^n$ and $\varphi(x) = (x^\top,1)^\top$, $C_q'(H)$ is a matrix composed of the expectations of $xx^\top$ and $x$.}

We can then define the Fenchel conjugate $C_q^\ast$ of $C_q$ as:
 $$C_q^\ast(\Sigma) = \sup_{H \in \mathbb{H}_d} \ \ \tr [ H \Sigma] - C_q(H).$$
  The domain of $C_q^\ast$ is then exactly the set of attainable moments (denoted $\mathcal{K}$ in \mysec{sos}), and the moment $\Sigma(H) = C_q'(H)$ is
 exactly the maximizer in 
$$
\sup_{ \Sigma \in \mathbb{H}_d } \ \tr [ H \Sigma ] - C^\ast_q(\Sigma).
$$
Note that in the future approximations of $C_q^\ast$ or $C_q$, there is both an approximation of the value \emph{and} potentially of the domain.

\paragraph{Estimation.} Given some data $x_1,\dots,x_n \in \X$, we can form the empirical moment $  \widehat{\Sigma} = \frac{1}{n} \sum_{i=1}^n \varphi(x_i) \varphi(x_i)^\ast$, and estimate $H \in \mathbb{H}_d$ by minimizing $D(p\|q)$ such that $\Sigma_p = \widehat{\Sigma}$. For $f(t) = t \log t - t + 1$, this is exactly maximum entropy estimation, which is classsicaly equivalent to finding the exponential family distributions with feature $x \mapsto \varphi(x) \varphi^\ast(x)$ and matching moment. This happens to be true for all $f$-divergences, that is, the optimal distribution $p$ is exactly $p = p(\cdot | H)$ for $H$ maximizing $\tr [ H \widehat{\Sigma} ] - C_q(H)$, and with matching moments. Note however that the formulation as the minimum (right) Kullback-Leibler divergence does not readily generalize beyond the Shannon entropy.

\section{Relaxed $f$-partition function}
\label{sec:partition}

Given the sequence of lower bounds  on the $f$-divergence $D^{\rm OPT} \geqslant D^{\rm SOS} \geqslant D^{\rm QT}
\geqslant \widetilde{D}^{\rm QT}_{\max}\geqslant \widetilde{D}^{\rm QT}_{\rm standard}  $, we get a sequence of upper-bounds for  $C_q(H)$ defined in \eq{cH}  for any Hermitian matrix $H \in \H_d$. {Since $q$ {(which is now assumed to sum to one)} only appears through 
$B = \Sigma_q$, we define our upper-bound as the maximal potential $C_q(H)$ for all distributions $q$ such that $\Sigma_q = B$, that is,}
{$$
C_B^{\rm OPT}(H)  = \sup_{q \in \mathcal{P}(\X)} C_q(H)  \mbox{ such that } \Sigma_q = B.
$$
We can now use the definition of $C_q(H)$ from \eq{cH} to get:
\BEA
\nonumber C_B^{\rm OPT}(H) 
&= &  \sup_{p,q \in \mathcal{P}(\X)} \ \int_\X   \varphi(x)^\ast H \varphi(x)   dp(x) - D(p\|q) \mbox{ such that } \Sigma_q = B \\
\nonumber & = &  \sup_{A \in \mathcal{C}, \ \tr[A]=1} \sup_{p,q \in \mathcal{P}(\X)} \ \int_\X   \varphi(x)^\ast H \varphi(x)   dp(x) - D(p\|q) \mbox{ such that } \Sigma_q = B \mbox{ and }\Sigma_p = A\\
\label{eq:CQH}  & = &        \sup_{ A \in \H_d  } \ \ \tr [A H ] - D^{\rm OPT}(A \| B)
  \ \mbox{ such that } \tr [ A ] = 1, \mbox{ by definition of } D^{\rm OPT}.
  \EEA
  Note that the constraints that $A \in \mathcal{V}$ and $A \succcurlyeq 0 $ are implied
  by the finiteness of $D^{\rm OPT}(A \| B)$ (this will not be the case for the other relaxations).}
  
  We use the variational representation of $D^{\rm OPT}(A \| B)$ in \eq{DSOS},  to get, using the Lagrange multiplier $\rho$ for the constraint $\tr [A] = 1$:
  \BEA
  \notag
  & &  C_B^{\rm OPT}(H)  \\
   \notag& = &  \!\!  \sup_{A\in \H_d  } \   \inf_{M,N \in \H_d } \ \  \tr [ A H ] - \tr[M A] - \tr[N B ] \  \mbox{ such that } \ \forall r \geqslant 0 , \ \Gamma( rM + N) \leqslant  f(r)   \mbox{ and } \tr [ A ] = 1
\\
   \notag & = &  \!\!   \sup_{A  \in \H_d  } \   \inf_{M,N\in \H_d ,\ \rho \in \rb }  \tr [ A H ] - \tr[M A] - \tr[N B ]  - \rho \big( \tr [A]-1 )  \ \mbox{ such that } \  \forall r \geqslant 0 , \ \Gamma( rM + N) \leqslant  f(r)  .
\EEA
{The convex set $\{(M,N) \in \H_d \times \H_d, \ \forall r \geqslant 0, \Gamma( rM + N) \leqslant  f(r) \}$ has non empty interior since $(0,-\idm)$ is in the interior; thus strong duality holds \cite[Section~8.6]{luenberger1997optimization}, and we can swap infimum and supremum. Taking the supremum with respect to $A$ leads to the equality constraint $H - M - \rho \idm = 0$, and thus}
\BEA
  \label{eq:Cqopt}  C_B^{\rm OPT}(H) 
  & = &    \!\!\inf_{N\in \H_d ,\ \rho \in \rb  } \ \    \rho   - \tr[N B ]  \mbox{ such that } \forall r \geqslant 0 , \ \Gamma( rH   + N) \leqslant  f(r)  + \rho r.
  \EEA
{By replacing $N$ by $\bar{N}-\rho \idm$, and then optimizing in closed form with respect to $\rho$, we get:}
  \BEA
  \notag  C_B^{\rm OPT}(H)  & = &   { \!\!\inf_{\bar{N}\in \H_d ,\ \rho \in \rb  } \ \    2 \rho   - \tr[\bar{N} B ]  \mbox{ such that } \forall r \geqslant 0 , \ \Gamma( rH   + \bar{N}) \leqslant  f(r)  + \rho (r+1)}
\\
  \label{eq:Cqoptbis} & = &   { \!\!\inf_{\bar{N}\in \H_d   } \ \    2 \sup_{r \geqslant 0} \frac{\Gamma(rH+\bar{N})-f(r)}{r+1}   - \tr[\bar{N} B ]= \inf_{N\in \H_d   } 2 G(H,\bar{N}) - \tr[\bar{N}B]  .}  
    \EEA
    This corresponds   to the formulation in \eq{www}. {This can now be solved with Kelley's method, like described in \mysec{sos-relax}.}
 
 \paragraph{Recovering the optimal moment matrix $\Sigma_p = A$.} {Given the solution $\bar{N}$ of \eq{Cqoptbis}, we get the optimal $\rho$ and $N$ of \eq{Cqopt} as $\rho = \sup_{r \geqslant 0} \frac{\Gamma(rH+\bar{N})-f(r)}{r+1}$
 and $N = \bar{N} - \rho \idm$, and $M = H - \rho \idm$. Optimality conditions for \eq{Cqoptbis} lead to
 $B = \int_{\X \times \rb_+} \frac{1}{r+1} \varphi(x)\varphi(x)^\ast d\nu(x,r)$, where $\nu$ is a probability distribution supported on the maximizers of $\frac{\varphi(x)^\ast (rM+N) \varphi(x)-f(r)}{r+1} $.
 Since $\tr[B] = 1$, we have $\int_{\X \times \rb_+} \frac{1}{r+1}   d\nu(x,r) = 1$. The optimal $A$ is 
 $A = \int_{\X \times \rb_+} \frac{r}{r+1} \varphi(x)\varphi(x)^\ast d\nu(x,r)$. We then have:
 $ \tr[ AM ] + \tr [NB] = \int_{\X \times \rb_+} \frac{f(r)}{r+1}   d\nu(x,r)  = D^{\rm OPT}(A\|B)$.
 
 }

    \paragraph{Tightness.} In this paper, we focus on the computation of upper-bounds of the partition function. The study of the approximation capabilities when the feature vector grows is left for future work. In particular, it would be interesting to compare to other convex upper-bounds on the log partition functions such as the ``tree-reweighted representation'' framework~\cite{wainwright2008graphical}.

We now consider computable relaxations, first based on SOS in \mysec{sos-part}, then on quantum information divergences in \mysec{qt-part}.

\subsection{Sum-of-squares relaxation}
\label{sec:sos-part}
  We get the SOS relaxation where $\Gamma$ is replaced by $\widehat{\Gamma}$ in \eq{Cqopt} and \eq{Cqoptbis}:
 \BEAS
  C_B^{\rm SOS}  (H)  & = & 
    \inf_{N \in \H_d,\ \rho \in \rb  } \ \    \rho   - \tr[N B ]  \mbox{ such that } \forall r \geqslant 0 , \   ( f(r)  + \rho r ) \idm - rH   - N \in \textcolor{red}{\widehat{\mathcal{C}}^\ast}  \\
  & = &   { \!\!\inf_{\bar{N}\in \H_d  } \ \    2 \sup_{r \geqslant 0} \frac{\widehat{\Gamma}(rH+\bar{N})-f(r)}{r+1}   - \tr[\bar{N} B ]  = \inf_{\bar{N}\in \H_d  } 2 \widehat{G}(H,\bar{N}) - \tr[\bar{N}B]}.
      \EEAS
This is now approximable in polynomial time and is an upper bound on $C^{\rm OPT}(H)$.  

{To obtain the optimal $A$ from an optimal $N$,  optimality conditions lead to  a probability measure $\nu$ on $\widehat{\mathcal{K}} \times \rb_+$, corresponding to the maximizers of $\sup_{r \geqslant 0} \sup_{\Sigma \in \widehat{\mathcal{K}} } \frac{\tr[\Sigma (rH+\bar{N})]-f(r)}{r+1} $ and
$A = \int_{\X \times \rb_+} \frac{r}{r+1} \Sigma d\nu(\Sigma,r)$.}

    \paragraph{Algorithms.} 
   {We can use the same technique as for computing $D^{\rm SOS}(A\| B)$ and use Kelley's method. It can be initialized by considering the spectral relaxation detailed below.}

     \subsection{Quantum relaxation}
\label{sec:qt-part}

 We now consider the quantum relaxation instead of the sum-of-squares relaxation, for $H \in \H_d$ (the constraint that $A \succcurlyeq 0$ is automatically satisfied but not the one that $A \in \mathcal{V}$), using convex duality:
  $$
 C_B^{\rm QT}(H)   = 
   \sup_{A  \in \mathcal{V} } \ \ \tr [A H ] - D^{\rm QT}(A \| B)
   \ \mbox{ such that } \tr [ A   ] = 1.
   $$
{In order to derive estimation algorithms, we  also propose a formulation based on the unconstrained formulation of $D^{\rm QT}$ from \eq{GMNQT}, and introducing a Lagrange multipler $Y \in \mathcal{V}^\perp$ for the constraint 
$A  \in \mathcal{V} $,  as (assuming $\tr[B]=1$):
   \BEQ
\label{eq:CQOPT}   C_B^{\rm QT}(H)   = \inf_{\bar{N},V \in \mathbb{H}_d,Y \in \mathcal{V}^\perp} - \tr[\bar{N}B]+ 2 \sup_{ r \geqslant 0} \frac{ \lambda_{\max}(r H+ rY + \bar{N} - f(r) V)}{r+1} \mbox{ such that } \ V \succcurlyeq 0, \ V  - \idm \in \mathcal{V}^\perp .
   \EEQ
   We can recover the optimal $A$ (which may not be in $\mathcal{V}$) like for the other relaxations.}

\paragraph{Spectral relaxation.} {In order to obtain approximate closed-form expressions for initialization, we could consider the spectral relaxation with $V=\idm$, and $A$ not constrained to be in $\mathcal{V}$, leading to:
$$
  \sup_{A  \in \H_d^+ } \ \ \tr [A H ] -  \tr\big[ B f(B^{-1/2} A B^{-1/2}) \big]   \ \mbox{ such that } \tr [ A   ] = 1,
   $$ 
   which cannot be solved in closed form in general. We can also consider the standard quantum relaxation, which does not lead to a closed form expression, except for the function $f(t) = t \log t - t + 1$, where we need to solve 
\BEQ
\label{eq:spectrall}  \sup_{A  \in \H_d^+ } \ \ \tr [A H ] -  \tr[ A \log A - A \log B - A + B]   \ \mbox{ such that } \tr [ A   ] = 1,
   \EEQ
   which is equal to $\log \tr \exp( H + \log B)$, with $A = \frac{ \exp( H + \log B) }{\tr  \exp( H + \log B)}$. The resulting function of $B$ is concave, its gradient is an initializer for $- {N}$, and it can be computed from the Jacobians of the exponential and logarithm maps. The last expression is not invariant to the addition of an element of $\mathcal{V}^\perp$ to $H$ (while it should). Following~\cite[Appendix B]{bach2022exponential}, we can make it invariant by projecting $H$ onto $\mathcal{V}$.
      }

\section{Experiments}
\label{sec:exp}

In this section,\footnote{{Matlab code to reproduce all experiments can be downloaded from \url{www.di.ens.fr/~fbach/fdiv_quantum_var.zip}.}} we illustrate our various relaxations and algorithms presented in earlier sections. We illustrate our results with the function $f(t) = t \log t - t + 1$, and thus the estimation of relative Shannon entropies and log-partition functions. {We focus on two particular examples, $[0,1]^n$ with trigonometric polynomials, and $\{-1,1\}^n$ with regular polynomials, as they correspond to generic multivariate continuous and discrete situations. In the discrete case of $\{-1,1\}^n$, exact computations have an exponential complexity in $n$, which is unavoidable~\cite{valiant1979complexity}. In contrast, the approximation methods provided in this paper have polynomial time in $n$, though with an exponent that grows fast with the maximal cardinality size in the feature $\varphi$. The continuous case creates extra difficulties depending on the order of polynomials that we consider (with the need to use high-order quadrature formulas~\cite{gautschi2011numerical} to compute quantities exactly). }

{We first consider in \mysec{expentropies} computing relative entropies and compare there all relaxations, from the more costly ones based on sums-of-squares (SOS) to quantum-based ones (QT) which rely on faster spectral computations. Given that the quantum-based ones have a very similar performance a much lower computational cost, we only consider these in \mysec{expvarinf} where we compute log-partition functions.}

\subsection{Computing relative entropies}
\label{sec:expentropies}

{We consider three classical situations, trigonometric polynomials on $[0,1]$, where the SOS relaxation is tight (and thus $D^{\rm OPT} = D^{\rm SOS}$), as well as trigonometric polynomials on $[0,1]^n$, and polynomials on $\{-1,1\}^n$.}

\paragraph{Trigonometric polynomials on $\X = [0,1]$.}
We consider $q$ the uniform distribution on $[0,1]$ (with density~$1$ with respect to the Lebesgue measure), and $p$ with density $ \frac{8}{\pi} \sqrt{x(1-x)}$. We have the following moments:
\BEAS
\int_{0}^1 e^{2 i\pi \omega x} dq(x) &= & 1 \ \mbox{ if } \omega = 0, \mbox{ and } 0 \mbox{ otherwise}, \\
\int_{0}^1 e^{2 i\pi \omega x} dp(x) & = & \frac{2 (-1)^\omega}{\omega  \pi} J_1( \omega  \pi) \mbox{ for } \omega \neq 0,
\EEAS
where $J_1$ is the Bessel function of the first kind, as well as the relative entropy $D(p\|q) \approx 0.0484$, {which can be approximated with high precision with quadrature formulas~\cite{gautschi2011numerical}.}

We consider $\omega \in \{-r,\dots,r\}$, and compute the various bounds:  OPT, SOS (which are equal), and QT (together with a version only optimized over diagonal $V$), and the old version of QT from~\cite{bach2022information} (where we only learn diagonal matrices $V$).

\begin{figure}
\begin{center}
\includegraphics[width=6.5cm]{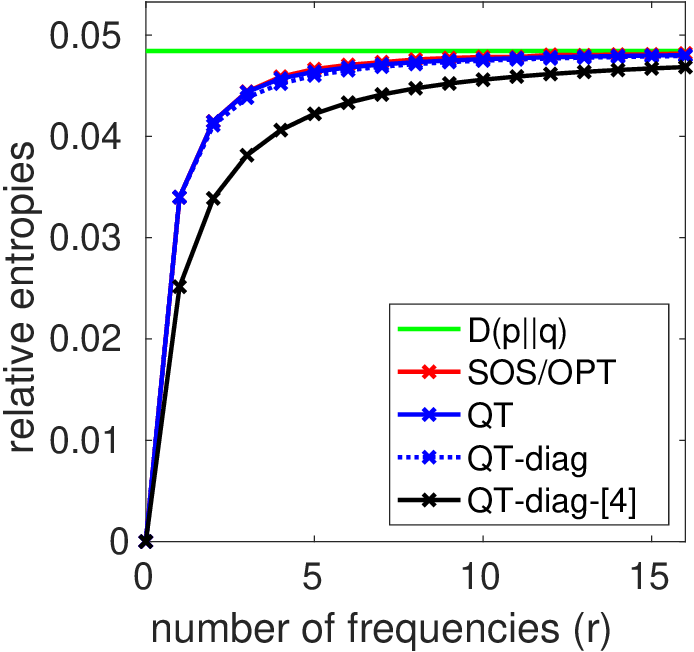}
\hspace*{1.5cm} 
\includegraphics[width=6.5cm]{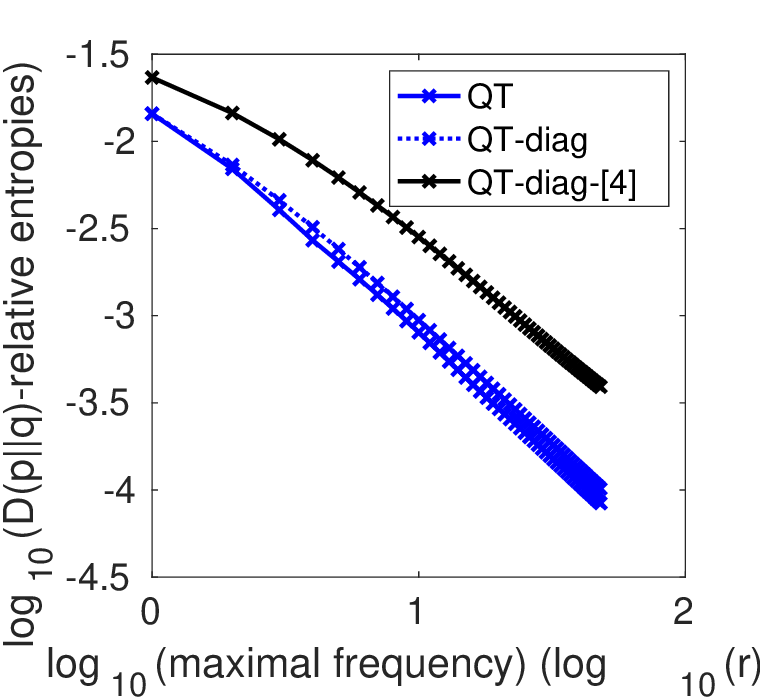}
\end{center}

\vspace*{-.3cm}

\caption{Comparison of relative entropy estimates for several numbers of frequencies, showing the effect of varying $r$.
Left: $D(p\|q)$ is the exact value, with moment-based approximations OPT/SOS (here equal), and the ones based on quantum information divergences, with full metric learning (QT) or diagonal metric learning (QT-diag), with also the use of the weaker quantum divergences proposed by~\cite{bach2022information}. {In several cases the red curve associated with OPT/SOS bounds is not visible as it coincides with the blue curve of QT bounds. Right: same experiment without OPT/SOS, with deviation to $f$-divergence reported in a log-log plot.}}
\label{fig:entropy_1d}
\end{figure}

We see in the left plot of \myfig{entropy_1d} that the optimal/SOS bound is numerically identical to the full quantum bound, and close to the one with diagonal $V$, but with a strong improvement over the bound from~\cite{bach2022information}. {For the SOS relaxations, results are obtained using Kelley's method described in \mysec{sos-relax}}.

{In the right plot of \myfig{entropy_1d}, we only compute the spectral relaxations, for significantly larger $r$, showing that, as $r$ grows, we get a tighter approximation of $D(p\|q)$ for all methods.}

\paragraph{Trigonometric polynomial on $\X = [0,1]^n$.}
We consider $x_1$ uniform on $[0,1]$ and $x_{i+1} = \lfloor x_i + \eta_{i+1} \rfloor$, where $\eta_{i+1}$ is uniform on $[-\rho/2,\rho/2]$. When $\rho=0$, all $x_i$'s are equal almost surely, while when $\rho=1$,   all $x_i$'s are independent and uniform {(we use $\rho = 3/10$ in our simulations)}.

We can then compute the Kullback-Leibler divergence to the uniform distribution by noticing that the sequence $(x_i)$ forms a Markov chain, so that (using classical entropy decomposition results for tree-structured graphical models~\cite{wainwright2008graphical}):
\BEAS
D(p\|q) & = & \int_{[0,1]^n} p(x) \log  {p(x)} dx  \\
& = &  \sum_{i=1}^{n-1}  \int_{[0,1]^2} p(x_i,x_{i+1}) \log  { p(x_i,x_{i+1}) } dx_i dx_{i+1}
- \sum_{i=2}^{n-1}  \int_{[0,1]} p(x_i) \log  { p(x_i) } dx_i     = (n-1) \log\frac{1}{\rho}.
\EEAS

We can also get all Fourier moments by introducing the $n \times (n-1)$ $\{0,1\}$-valued matrix $M$ such that $ x_i = (M \eta)_i + x_1$ for $i \in \{2,\dots,n\}$. We then have 
$$
\E [   e^{2i\pi \omega^\top x} ] = 1_{\omega^\top 1 = 0} \cdot \prod_{k=1}^{n-1} \frac{ \sin \big[  (M^\top \omega)_k \pi \rho\big] }{(M^\top \omega)_k \pi \rho }.
$$

\begin{figure}
\begin{center}
\includegraphics[width=6.5cm]{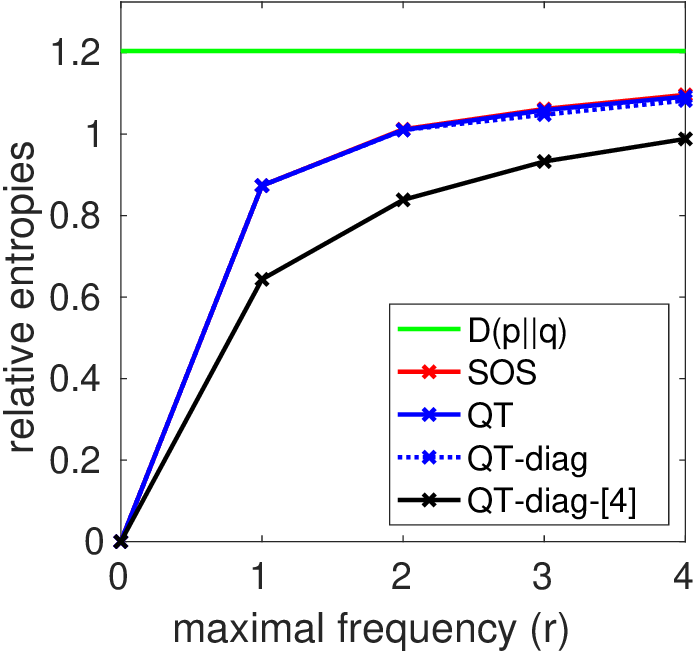}
\hspace*{1.5cm} 
\includegraphics[width=6.5cm]{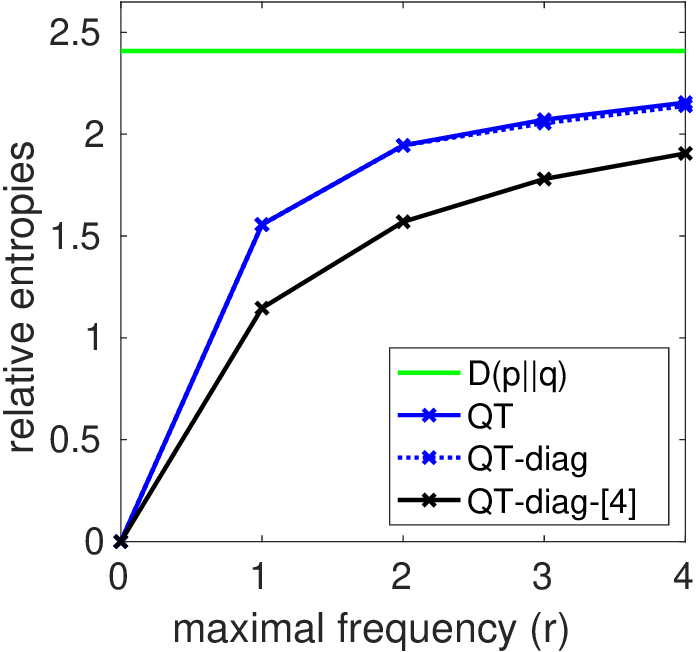}
\end{center}

\vspace*{-.3cm}

\caption{Comparison of relative entropy estimates for several numbers of maximal marginal frequency $r$, for $n=2$ (left) and $n=3$ (right). $D(p\|q)$ is the exact value, with moment SOS-based approximations (only plotted in the left plot, {and for which we only compute the cost function of \eq{GMN} with the optimal quantum solution from \eq{VV}, which is close to optimal for \eq{GMN}}), and the ones based on quantum information divergences, with full metric learning (QT) or diagonal metric learning (QT-diag), with also the use of the weaker quantum divergences done by~\cite{bach2022information}.
\label{fig:entropy_multipled}}
\end{figure}

In order to estimate entropies, we consider $\|\omega \|_\infty\leqslant r$. See \myfig{entropy_multipled}, {where we can draw similar conclusions as for $n=1$.}

 \paragraph{Polynomials on $\X = \{-1,1\}^n$}
We consider the  task of estimating entropies from moments on a simple example, where we consider $x_1$ uniform on $\{-1,1\}$ and $x_{i+1} = x_i \eta_{i+1} $, where $\eta_{i+1} \in \{-1,1\}$ is independent and equal to $1$ with probability $1-\rho/2$, and $-1$ otherwise. When $\rho=0$, all $x_i$'s are equal almost surely, while when $\rho=1$,  all $x_i$'s are independent and uniform {(we use $\rho = 1/2$ in our simulations)}.

\begin{figure}
\begin{center}
\includegraphics[width=7cm]{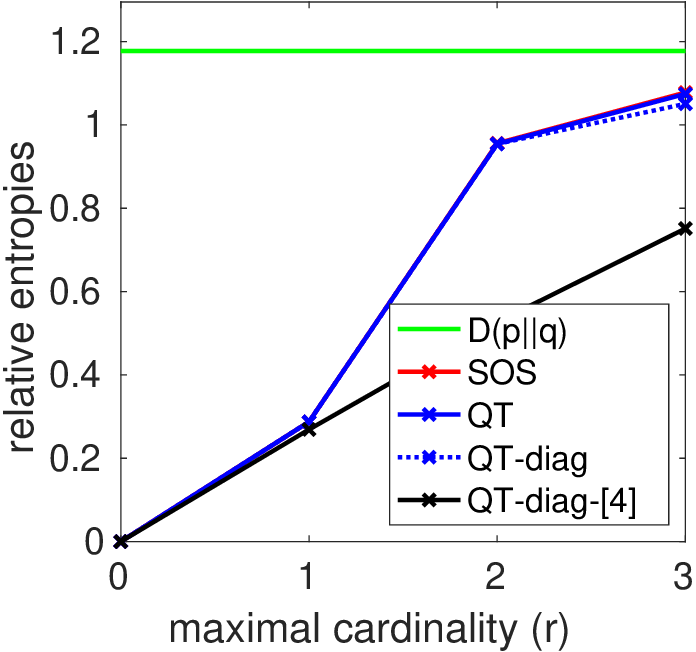}
\hspace*{2cm} 
\includegraphics[width=7cm]{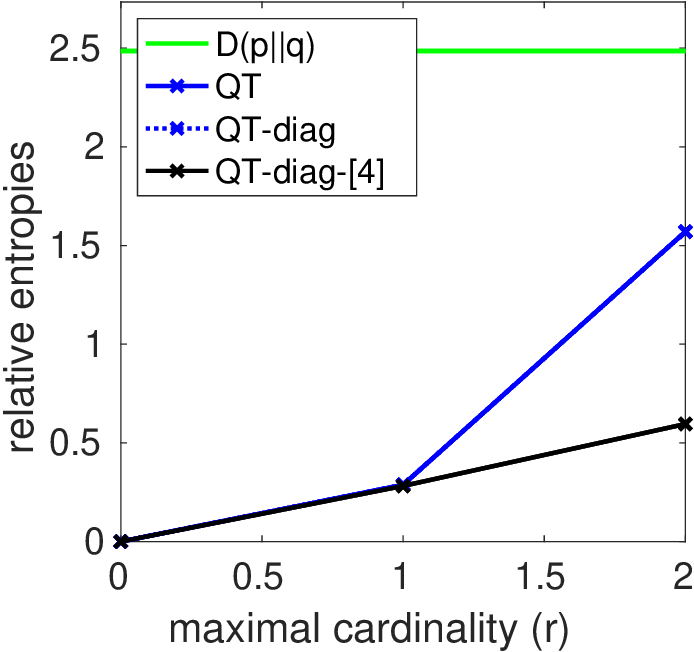}
\end{center}

\vspace*{-.3cm}

\caption{Comparison of relative entropy estimates for several numbers of frequencies for all subsets of cardinality less than $r$, for $n=10$ and $n=20$. $D(p\|q)$ is the exact value, with moment SOS-based approximations, and the ones based on quantum information divergences, with full metric learning (QT) or diagonal metric learning (QT-diag), with also the use of the weaker quantum divergences done by~\cite{bach2022information}. 
\label{fig:entropy_boolean}}
\end{figure}

We can then compute the Kullback-Leibler divergence to the uniform distribution in the same way as for data in $[-1,1]^n$, leading to
$
D(p\|q) = (n-1) \big[  ( 1 - \frac{\rho}{2} ) \log( 2 - \rho) + \frac{\rho}{2} \log \rho \big].
$
We can also get all Fourier moments as
$
\E \big[ \prod_{i\in A} x_i \big] = ( 1 - (-1)^{|A|}) \prod_{i=1}^{n-1} ( 1  - \rho)^{1- (-1)^{(M^\top 1_A)i}}.
$
In order to estimate entropies, we consider subsets of cardinality less then $r$. See \myfig{entropy_boolean}, for $n=10$ and $n=20$, {where we can draw similar conclusions as for $n=1$.}

\subsection{Computing log-partition functions}
\label{sec:expvarinf}

{We now compare algorithms to upper-bound log-partition functions, by only focusing on the more efficient quantum relaxations. We do so for trigonometric polynomials on $[0,1]$.}

\paragraph{Log-partition functions  on $\X = [0,1]$.}
We consider $h(x) = \cos(4\pi x)$, with $\log   \int_{0}^1 e^{\cos(4\pi x)} dx = \log I_0(1) \approx 0.2359$, where we use the same feature map $\varphi : [0,1] \to \cb^{2r+1}$ as before, which enables us to write $h(x) = \varphi(x)^\ast H \varphi(x)$ for some Hermitian matrix~$H$. {The matrix $H$ is not unique, and, following~\cite[Appendix B]{bach2022exponential}, we consider the spectral relaxation from \eq{spectrall},
with $H$ orthogonally projected on the set $\mathcal{V}$ of Toeplitz matrices. This spectral relaxation was considered in \cite{bach2022information} but in a positive definite kernel context where the projection onto Toeplitz matrices is not applicable. Such a projection is crucial to obtain a meaningful result.}

  We then compute the approximation $C_B^{\rm QT}$  {by using Kelley's method to solve \eq{CQOPT}, and we also report results without the optimization with respect to $V$, with an almost identical curve (showing that the benefits of metric learning are here marginal). While for small $r$, the new quantum relaxation improves over the bound adapted from $\cite{bach2022information}$, it does not for larger $r$.}

\begin{figure}
\begin{center}
\includegraphics[width=7cm]{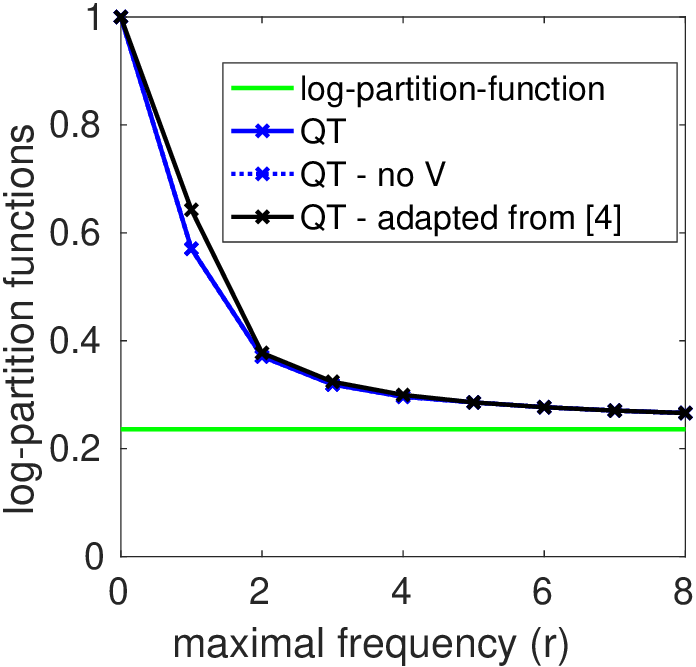}  
\end{center}

\vspace*{-.3cm}

\caption{Comparison of log-partition function estimates for several numbers of frequencies for $n=1$. See text for details. \label{fig:partition_1d}}
\end{figure}

\section{Conclusion}
\label{sec:conclusion}
In this paper, we have proposed to combine tools from information theory, both classical such as $f$-divergences, and more recent, such as quantum information divergences, with sum-of-squares optimization.  This leads to several relaxations of $f$-divergences based on sum-of-squares relaxations or quantum information divergences, together with efficient estimation algorithms for the tasks of divergence estimation from moments and the computation of log-partition functions. {These relaxations are summarized in Table~\ref{tab:summary}.} While the relaxation  based on sums-of-squares ($D^{\rm SOS}$) is strictly superior, it is only mildly so in our experiments compared to the one based on quantum divergences ($D^{\rm QT})$, while being more costly to compute. This thus highlights the benefits of the quantum relaxation.

\begin{table}
\begin{center}
{\begin{tabular}{|l|c|l|}
\hline
$\textcolor{white}{\Big|} D^{\rm OPT}$  & \mysec{exact} & Not computable \\
\hline$\textcolor{white}{\Big|} D^{\rm SOS} $ & \mysec{sos-relax} & Computable with SOS \\
\hline
$ \textcolor{white}{\Big|} D^{\rm QT}$ & \mysec{qt-relax} & Computable with spectral method + single SDP\\
\hline
$\textcolor{white}{\Big|} \widetilde{D}^{\rm QT}_{\max}$ & \mysec{qt-relax} & Computable with spectral method \\
\hline
$\textcolor{white}{\Big|} \widetilde{D}^{\rm QT}_{\rm standard}$ &  \cite{bach2022information} & Computable with spectral method \\
\hline
\end{tabular}}
\end{center}

\vspace*{-.4cm}

{\caption{Summary of relaxations ordered by strength (function values and domains), from stronger (top) to weaker (bottom). \label{tab:summary}}}
\end{table}

This quantum information relaxation takes its roots in earlier work \cite{bach2022information} {with significant improvements: (a) the use of a tighter quantum divergence (the ``maximal'' one rather than the ``standard'' one), (b) the introduction of the optimal lower bound, (c) taking into account the particular geometries of feature vectors using sum-of-square techniques that improve over spectral relaxations, and (d) the proposal of generic optimization algorithms.} Several avenues are worth exploring: (a) check if the new notion of relative entropy with maximal divergence preserves properties from \cite{bach2022information}, in particular its use in probabilistic modelling and within graphical models, (b) potentially extend the positive definite kernel motivation that allows infinite-dimensional moments, {along the lines of~\cite{giraldo2014measures} which explored this connection for Renyi entropies}, (c)  obtain convergence rates for entropies and log-partition function estimation to go with our encouraging empirical results, (d) develop algorithms to deal with larger scale problems using approximation techniques from kernel methods~\cite{boutsidis2009improved,rudi2015less}.

\subsubsection*{Acknowledgements}
The author would like to thank Omar Fawzi for discussions related to quantum information divergences, Adrien Taylor and Justin Carpentier for discussions on optimization algorithms, as well as David Holzm\"{u}ller for providing clarifying comments. The comments of the anonymous reviewers were greatly appreciated. We   acknowledge support from the French government under the management of the Agence Nationale de la Recherche as part of the ``Investissements d’avenir'' program, reference
ANR-19-P3IA-0001 (PRAIRIE 3IA Institute), as well as from the European Research Council
(grant SEQUOIA 724063).
 
 \appendix

 \section{Newton method for computing $F$}
 \label{app:newton}
 
 {In order to compute $F(v,w) = \sup_{r \geqslant 0} \frac{ r  v + w - f(r)}{r+1}$ in \mysec{varf}, we assume $(v,w)\in \rb^2$ fixed and define $\varphi(r) = \frac{ r  v + w - f(r)}{r+1}$, which is twice differentiable, and so that, $\varphi'(r) = \frac{1}{(r+1)^2} \big[ v-w - \big( (r+1)f'(r) - f(r) \big) \big]$. Since the function $r \mapsto  (r+1)f'(r) - f(r) $ has derivative $r \mapsto (r+1)f''(r)$, it is  strictly positive. Thus the derivative $\varphi'$ thus has at most one zero, and we aim to solve the equation $  (r+1)f'(r) - f(r)  = v - w$.
 
We now focus on the special case  $f(t) = t \log t - t + 1$, where
$ \psi(r) = (r+1)f'(r) - f(r)  = r - 1 + \log r$, and has full range on $\rb_+$ so that the equation above has a unique solution. In order to compute its solution, for $v-w \geqslant 0$, we iterate Newton's method~\cite[Section~4.8]{gautschi2011numerical} $r \leftarrow r + \frac{v-w - \psi(r)}{\psi'(r) }
= \frac{2 - \log(r) + v-w}{ 1 + 1/r}$ for 5 iterations to reach machine precision, while for $v-w \leqslant 0$
we iterate  Newton's method on the logarithm of $r$
$  \log r  \leftarrow \log r +   \frac{v-w - \psi(r)}{r \psi'(r) }
 =    \frac{ 1 + v-w  + ( \log r -1) r}{ 1 + r} $ for 5 iterations.}

 \section{Decomposition of operator-convex functions}
 \label{app:op-cvx}
 
 We have the following particular cases from \mysec{f-review}.
 
\BIT
\item $\alpha$-divergences: $  f(t) =\frac{1}{\alpha(\alpha-1)} \big[   t^\alpha - \alpha t + (\alpha-1) \big] =
 \frac{1}{\alpha} \frac{\sin (\alpha-1)\pi}{(\alpha-1)\pi} (t-1)^2 \int_0^{+\infty} \frac{1}{t+\lambda} \frac{ \lambda^\alpha d\lambda}{(1+\lambda)^2}
$
for  $\alpha \in (-1,2)$. Other representations exist for $\alpha=-1$ and $\alpha=2$ (see below), but other cases are not operator-convex.
\item KL divergence ($\alpha=1)$: $f (t) = t \log t -t + 1 = \int_0^{+\infty} \frac{(t-1)^2}{t+\lambda} \frac{ \lambda d\lambda}{(\lambda+1)^2}$.
\item Rerverse KL divergence  ($\alpha=0)$: $  f(t) = - \log t + t - 1     = \int_0^{+\infty} \frac{(t-1)^2}{t+\lambda} \frac{   d\lambda}{(\lambda+1)^2}$.

\item Pearson $\chi^2$ divergence ($\alpha=2$): $f(t) = \frac{1}{2} (t-1)^2$ is operator convex.

\item Reverse pearson $\chi^2$ divergence  ($\alpha=-1$): $f(t) = \frac{1}{2} \big( \frac{1}{t} + t \big) - 1 = \frac{1}{2} \frac{(t-1)^2}{t}$ is operator convex, with  $d\nu(\lambda)$ proportional to a Dirac at $\lambda = 0$.

\item Le Cam distance: $f(t)  = \frac{(t-1)^2}{t+1}$ is operator convex with $d\nu(\lambda)$ proportional to a Dirac at $\lambda = 1$.

\item Jensen-Shannon divergence: $f(t) =2 t \log \frac{2t}{t+1} + 2 \log \frac{2}{t+1} =2 t \log t - 2(t+1) \log(t+1) + 2(t+1) \log 2
=   2t \log t - 4 \frac{t+1}{2}\log\frac{t+1}{2}  $ is operator convex, as it can be written
$  f(t) = 2(t-1)^2 \int_0^{+\infty} \big( \frac{1}{t+\lambda} - \frac{1}{t+1+2\lambda}
\big)  \frac{\lambda d\lambda}{(1+\lambda)^2} $, which leads to
$  f(t) =2 (t-1)^2 \int_0^{+\infty}   \frac{1}{t+\lambda} \frac{\lambda d\lambda}{(1+\lambda)^2} - 
2 (t-1)^2  \int_1^{+\infty}   \frac{1}{t+\lambda} \frac{(\lambda-1) d\lambda}{(1+\lambda)^2}$.  \EIT

\section{Proofs of  Lemma~\ref{lemma:ABL} and  Lemma~\ref{lemma:ABLV} }
\label{app:lemma}
In this section, we prove Lemma~\ref{lemma:ABL} from \mysec{qt-relax}, taken from \cite[Section 9.1]{matsumoto2015new} and shown in here for completeness and the expression of the maximizers. We also prove the extension Lemma~\ref{lemma:ABLV}.

We start by the dual formulation:
\BEAS
 &  &  \sup_{M,N \in \H_d } \ 
 \tr [ MA ] + \tr [ NB] \  \mbox{ such that }  \  \forall r \geqslant 0, \   r M  + N \preccurlyeq f(r) \idm \\
 \notag & = &  \inf_{\Lambda  \  \textcolor{red}{\mathbb{H}_d^+}-{\rm valued \ measure \ on}\ \rb_+}    \int_0^{+\infty}\!
f(r) \tr \big[ d\Lambda(r)     ]    \  \mbox{ such that } \ \int_0^{+\infty}\!
  d\Lambda(r)      = B \mbox{ and } \int_0^{+\infty} \!
 r d\Lambda(r)   = A,
 \EEAS
 which can be solved in closed form.

Indeed, given the eigendecomposition $B^{-1/2} A B^{-1/2} = \sum_{i =1}^d \lambda_i u_i u_i^\ast$, we consider
$\Lambda = \sum_{i =1}^d B^{1/2}  u_i u_i^\ast B^{1/2}  \delta_{\lambda_i}$, where $\delta_{\lambda_i}$ is the Dirac measure at $\lambda_i$,  so that we get a feasible measure $\Lambda$, and an objective equal to 
$ \sum_{i=1}^d f(\lambda_i)  \tr \big[B^{1/2} V B^{1/2} u_i u_i^\ast \big] = \tr \big[ B^{1/2} V B^{1/2} f \big( B^{-1/2}AB^{-1/2} \big) \big]$. Thus the infimum is less than $ \tr \big[ B^{1/2} V B^{1/2} f \big( B^{-1/2}A B^{-1/2} \big) \big]$.

The other direction is a direct consequence of the operator Jensen's inequality~\cite{hansen1982jensen}: 
for any feasible measure~$\Lambda$ approached by an empirical measure $\ds\sum_{i =1}^m M_i \delta_{r_i}$, with $M_i \succcurlyeq 0$, we have $\ds \sum_{i =1}^n( M_i^{1/2} B^{-1/2} )^\ast ( M_i^{1/2} B^{-1/2} ) = \idm$, and thus
\BEAS
\int_{\rb_+} \! f(r) d\Lambda(r) &  =  & B^{1/2} \Big(\sum_{i =1}^m ( M_i^{1/2} B^{-1/2} )^\ast f(r_i\idm) ( M_i^{1/2} B^{-1/2} ) \Big) B^{1/2} \\
 & \succcurlyeq & B^{1/2} f \Big(\sum_{i =1}^m( M_i^{1/2} B^{-1/2} )^\ast (r_i \idm) ( M_i^{1/2} B^{-1/2} ) \Big)  B^{1/2}  =   B^{1/2}   f \big( B^{-1/2}A B^{-1/2} \big)  B^{1/2}.\EEAS
  The lower bound follows by letting the number $m$ of Diracs go to infinity to tightly approximate any feasible matrix $\Lambda$. 
  
  In order to obtain the minimizers $M$ and $N$, we simply notice that they are the gradients of the function $(A,B) \mapsto \tr \big[ B  f \big( B^{-1/2}A B^{-1/2} \big) \big]$ with respect to $A$ and $B$. We thus get, using gradients of spectral functions
\BEAS
M^\ast & = & \sum_{i,j=1}^d  \frac{f(\lambda_i)-f(\lambda_j)}{\lambda_i - \lambda_j} u_i^\top B u_j  \cdot B^{-1/2} u_i u_j^\ast B^{-1/2},
\EEAS
with the convention that for $\lambda_i = \lambda_j$, $ \frac{f(\lambda_i)-f(\lambda_j)}{\lambda_i - \lambda_j} = f'(\lambda_i)$. To obtain $N^\ast$, we simply consider the identity 
$\tr \big[ B  f \big( B^{-1/2}A B^{-1/2} \big) \big] = \tr \big[ A  g \big( A^{-1/2}B A^{-1/2} \big) \big]$, for $g(t) = t f(1/t)$, and use the corresponding formula.

\paragraph{Proof of  Lemma~\ref{lemma:ABLV}.}{We simply need to change slightly the proof of the lemma above, by replacing $\int_0^{+\infty}\!
f(r) \tr \big[ d\Lambda(r) \big] $ by $\int_0^{+\infty}\!
f(r) \tr \big[ V d\Lambda(r) \big]$ with all inequalities preserved because $V \succcurlyeq 0$.}

\section{Computing integrals}
\label{app:computingintegrals}
\label{app:integrals}

{A third task is related to $f$-divergences beyond computing the divergences themselves and the associated log-partition functions.} In this section, we consider the task of computing  $  \int_{\X} f^\ast ( h(x)) dq(x)$, where $q$ is a finite positive measure on $\X$, $f^\ast$ is the Fenchel conjugate of $f$, and $h: \X \to \rb$ an arbitrary function (such that the integral is finite). {The difference with computing log-partition functions in \eq{cqhi} is minor and we thus extend only a few results from \mysec{varinf} and \mysec{partition}, most without proofs as they follow the same lines as results for $f$-partition functions.}

For  $f(t) = t \log t - t + 1$, we have $f^\ast(u) =e^u - 1$, and we there aim at estimating integrals of exponential functions, a classical task in probabilistic modelling (see~\cite{wainwright2008graphical,murphy2012machine} and references therein), {which up to a logarithm is the same as computing the log-partition function; however, they are different for other functions $f$.}

This computational task can be classically related to $f$-divergences by Fenchel duality as we have:
$$
 \int_{\X} f^\ast ( h(x)) dq(x) = \sup_{p\  {\rm positive \ measure \ on}\  \X}\  \int_\X h(x) dp(x) - D(p\|q),
$$
{where the only difference with \eq{donsker} is that $p$ is not assumed to sum to one.}
Below, we show that for  functions $h(x)$ which are quadratic forms in $\varphi(x)$, we can replace $D(p\|q)$ by the lower-bound we just defined above, and obtain a computable upper bound of the integral.

We also have the representation corresponding to \eq{www}, that will be useful later:
\BEQ
\label{eq:Z} \int_\X f^\ast(h(x)) dq(x) =  \inf_{ w: \X \to \rb}     -  \int_\X w(x)  dq(x)  \  \mbox{ such that }\ \forall x \in \X, \forall r \geqslant 0, \  r h(x)  + w(x) \leqslant f(r).
\EEQ

\paragraph{Related work.}
There exist many ways of estimating integrals, in particular in  compact sets in small dimensions, where various quadrature rules, such as the trapezoidal or Simpson's rule, can be applied to compute integrals based on function evaluations, with well-defined convergence rates~\cite{simpson}. In higher dimensions, still based on function evaluations, Bayes-Hermite quadrature rules~\cite{o1991bayes}, and the related kernel quadrature rules~\cite{chen2010super,bach2012equivalence} come with precise convergence rates linking approximation error and number of function evaluations~\cite{bach2017equivalence}. An alternative in our context is Monte-Carlo integration from samples from $q$~\cite{robert1999monte}, with convergence rate in $O(1/\sqrt{n})$ from $n$ function evaluations.

In this paper, we follow~\cite{bertsimas2008approximating} and consider computing integrals given a specific knowledge of the integrand, here of the form $f^\ast(h(x))$, where $h$ is a known quadratic form in a feature vector $\varphi(x)$. While we also use a sum-of-squares approach as in~\cite{bertsimas2008approximating}, we rely on different tools (link with $f$-divergences and partition functions rather than integration by parts).

\paragraph{Relaxations.}
In order to compute integrals, we simply use the same technique but without the constraint that measures sum to one, that is, without the constraint that $\tr[A]=1$. Starting from \eq{Z}, we get, with $B = \Sigma_q$:
\BEAS
 \widetilde{C}_q(H) = \int_\X f^\ast\big(   \varphi(x)^\ast H \varphi(x)  \big) dq(x) 
 & = &    \sup_{ p\in \mathcal{M}_+(\mathcal{X})} \ \int_\X   \varphi(x)^\ast H \varphi(x)  dp(x) - D(p\|q) 
 \\
 & \leqslant  &  \sup_{ A \in \mathcal{C} } \  \tr[ HA]  - D^{\rm OPT}(A\|B)   =  \widetilde{C}_q^{\rm OPT} (H)  \\
 & = &   \inf_{N \in \H_d  } \ \    - \tr[N B ]  \mbox{ such that } \forall r \geqslant 0 , \ \Gamma( rH   + N) \leqslant  f(r).
\EEAS
Note that we only have an inequality here because we are not optimizing over $q$. We then get two computable relaxations by considering $D^{\rm SOS}(A\|B) $ and $D^{\rm QT}(A\|B) $ instead of $D^{\rm OPT}(A\|B) $, with the respective formulations:
\BEAS
 \widetilde{C}_q^{\rm SOS} (H) & = & \inf_{N \in \H_d  } \ \        - \tr[N B ]  \mbox{ such that } \forall r \geqslant 0 , \ f(r)   U - rH   - N \in \widehat{\mathcal{C}}^\ast   \\
 \widetilde{C}_q^{\rm QT} (H) & = & \sup_{A \in \H_d }  \ \inf_{V \succcurlyeq 0, \ V - \idm \in \mathcal{V}^\perp}  \ \tr [A H ] -   \tr \big[ B^{1/2} V B^{1/2}  f \big(B^{-1/2}A B^{-1/2} \big) \big].
\EEAS
Dual formulations and algorithms can then easily be derived.

\section{Dual formulations}
\label{app:varmin}

{In this appendix we present dual variational formulations to most of the formulations proposed in the main paper. We only consider the relaxations of $f$-divergences. Formulations for $f$-partition functions can be derived similarly.}

\paragraph{$f$-divergences.}
We can consider the Lagrangian dual of \eq{f-var}, by introducing a Lagrange multiplier $\lambda$ for the infinite-dimensional constraint $\forall x \in \X, \forall r \geqslant 0, \  r v(x)  + w(x) \leqslant f(r)$ in the form of a positive finite measure $\lambda$ on $  \X \times \rb_+$~\cite{jahn2020introduction}. We then obtain, {using strong duality for equality constraints~\cite[Section~8.6]{luenberger1997optimization}}:
\BEA
\notag\!\!\!\!\! D(p\|q) & = & {
\inf_{ \lambda\in  \mathcal{M}_+( \X \times \rb_+) }
  \sup_{ v,w : \X \to \rb }  \int_\X v(x) dp(x)  + \int_\X w(x) dq(x)   + 
  \int_\X \int_{\rb_+}  \big[ f(r) - rv(x) - w(x) \big] d\lambda(x,r)}
\\
\label{eq:Ddual} & = & \inf_{ \lambda\in  \mathcal{M}_+( \X \times \rb_+) } \ \int_\X \int_{\rb_+}   f(r)  d\lambda(x,r) \\[-.1cm]
\notag & & \hspace*{3.5cm} \mbox{ such that } \int_{\rb_+} \!d\lambda(\cdot,r)  = dq(\cdot)
\mbox{ and }   \int_{\rb_+} \! r d\lambda(\cdot,r)     = dp(\cdot),
\EEA
{with the two constraints resulting from the maximization with respect to $v$ and $w$.}

\paragraph{Optimal relaxation of $f$-divergences ($D^{\rm OPT}$).}

We can also formulate \eq{opty} as a minimization problem; we we can use Lagrangian duality, akin to \eq{Ddual}. This requires to introduce a Lagrange multiplier for the constraint 
$\forall r \geqslant 0, \ \Gamma(rM+N) \leqslant  f(r)$, which is equivalent to,
$ \forall r \geqslant 0, \ f(r)  -  rM-N \in \widehat{\mathcal{C}}^\ast$, which leads to a ${\mathcal{C}}$-valued finite measure on $\rb_+$~\cite{jahn2020introduction},
to get:
\BEA
  \notag \!\!\!\!&&D^{\rm OPT} ( A \| B ) \\
 \notag \!\!\!\!\!\!& = &  \inf_{\Lambda \   { \mathcal{C}}-{\rm valued \ measure \ on}\ \rb_+} \ \ \sup_{M,N \in \mathbb{H}_d} \ \ \tr [ AM ] + \tr [ BN] + \int_0^{+\infty}\!
\tr \big[ d\Lambda(r) ( f(r)   - rM - N ) \big] \\
\!\!\!\! \!\!& = &  \!\!\!\inf_{\Lambda  \   { \mathcal{C}}-{\rm valued \ measure \ on}\ \rb_+}    \int_0^{+\infty}\!
f(r) \tr \big[ d\Lambda(r)     ]    \  \mbox{ such that } \ \int_0^{+\infty}\!
  d\Lambda(r)      = B \mbox{ and } \int_0^{+\infty} \!
 r d\Lambda(r)   = A. \hspace*{1cm} 
 \label{eq:DOPTlambda}
\EEA

\paragraph{SOS relaxation of $f$-divergences ($D^{\rm SOS}$).}
 We can also get a formulation for \eq{ZZ}, akin to \eq{Ddual} and \eq{DOPTlambda}. This requires to introduce a Lagrange multiplier for the constraint $ \forall r \geqslant 0, \ f(r) U -  rM-N \in \widehat{\mathcal{C}}^\ast$, which is a $\widehat{\mathcal{C}}$-valued finite measure on $\rb_+$~\cite{jahn2020introduction},
to get:
\BEA
  \notag \!\!\!\!&&D^{\rm SOS} ( A \| B ) \\
 \notag \!\!\!\!\!\!& = &  \inf_{\Lambda \  \textcolor{red}{\widehat{\mathcal{C}}}-{\rm valued \ measure \ on}\ \rb_+} \ \ \sup_{M,N \in \mathbb{H}_d} \ \ \tr [ AM ] + \tr [ BN] + \int_0^{+\infty}\!
\tr \big[ d\Lambda(r) ( f(r)   - rM - N ) \big] \\
\!\!\!\! \!\!& = &  \!\!\!\inf_{\Lambda  \  \textcolor{red}{\widehat{\mathcal{C}}}-{\rm valued \ measure \ on}\ \rb_+}    \int_0^{+\infty}\!
f(r) \tr \big[ d\Lambda(r)     ]    \  \mbox{ such that } \ \int_0^{+\infty}\!
  d\Lambda(r)      = B \mbox{ and } \int_0^{+\infty} \!
 r d\Lambda(r)   = A, \hspace*{1cm} 
 \label{eq:DSOSlambda}
\EEA
{as maximizing out $M,N \in \mathbb{H}_d$ introduces linear constraints.}

\paragraph{Quantum relaxations of $f$-divergences ($D^{\rm SOS}$).}
 {\eq{QTdef} also has a primal-dual form as}
\BEA
 \label{eq:DQTPD}
 D^{\rm QT}(A\|B)  & = & \sup_{V \succcurlyeq 0, \ V  - \idm \in \mathcal{V}^\perp}  \ \
   \inf_{\Lambda  \   {\mathbb{H}_d^+}-{\rm valued \ measure \ on}\ \rb_+}    \int_0^{+\infty}\!
f(r) \tr \big[ d\Lambda(r) V   ]      \\
\nonumber & & \hspace*{4.5cm} \mbox{ such that }  \int_0^{+\infty}\!
  d\Lambda(r)      = B \mbox{ and } \int_0^{+\infty} \! r d\Lambda(r)      = A.
  \EEA
We also have a formulation akin to \eq{DSOSlambda}, that is, {adding a Lagrange multiplier $\Sigma \in \mathcal{V}$ in  \eq{DQTPD} for the constraint $\idm - V \in \mathcal{V}^\perp$:}
\BEAS D^{\rm QT}(A\|B) & =  & \inf_{\Sigma \in \mathcal{V}, \ \Lambda  \   {\H_d^+}-{\rm valued \ measure \ on}\ \rb_+}    \int_0^{+\infty}\!
f(r) \tr \big[ d\Lambda(r)  ] + \tr[ \Sigma]   \\
& &\hspace*{1.5cm}  \mbox{ such that }  \int_0^{+\infty}\!
  d\Lambda(r)      = B , \int_0^{+\infty}
 r d\Lambda(r)   = A,  \mbox{ and }  \int_0^{+\infty}\!
f(r)   d\Lambda(r) \preccurlyeq \Sigma, \EEAS
which shows the additional relaxation compared to $D^{\rm SOS}(A\|B)$, for which $\Lambda$ is a measure (almost everywhere) valued in $\mathcal{V}$, {while here it is only in $\mathbb{H}_d^+$.}

 \bibliography{quantum}

\textcolor{white}{.}
\thispagestyle{empty}

  \end{document}